\documentclass[reprint]{revtex4-1}

\usepackage[margin=1.0in]{geometry}

\usepackage{bbold}
\usepackage{amssymb}
\usepackage{graphics}
\usepackage{graphicx}

\usepackage{setspace}
\usepackage[font=singlespacing]{caption}

\usepackage{epstopdf}

\usepackage{amsmath}
\usepackage{amsthm}
\usepackage{amstext}
\usepackage{amscd}
\usepackage{epsfig}
\usepackage[mathscr]{eucal}
\usepackage{times}
\usepackage{srcltx}
\usepackage{shadow}
\usepackage{url}
\usepackage{color}
\usepackage[all]{xy}

\usepackage[most]{tcolorbox}

\usepackage{comment}
\usepackage{hyperref}

\usepackage{natbib}
\setcitestyle{authoryear,open={(},close={)}}

\begin{document}

\title{Evolution of specialized microbial cooperation in dynamic fluids}

\author{Gurdip Uppal, Dervis Can Vural}
\affiliation{Department of Physics, University of Notre Dame, Notre Dame, IN 46556}



\begin{abstract}
Here, we study the evolution of specialization using realistic computer simulations of bacteria that secrete two public goods in a dynamic fluid. Through this first principles approach, we find physical factors such as diffusion, flow patterns, and decay rates are as influential as fitness economics in governing the evolution of community structure, to the extent that when mechanical factors are taken into account, (1) Generalist communities can resist becoming specialists, despite the invasion fitness of specialization, (2) Generalist and specialists can both resist cheaters despite the invasion fitness of free-riding, (3) Multiple community structures can coexist despite the opposing force of competitive exclusion. Our results emphasize the role of spatial assortment and physical forces on niche partitioning and the evolution of diverse community structures. 
\end{abstract}

\maketitle

\section*{Introduction}

From subcellular structures to ecological communities, life is organized in compartments and modules performing specific tasks.
Organelles \citep{siegel1960hereditary,kutschera2005endosymbiosis}, single \citep{lewis2007persister} and multi-phenotype \citep{koufopanou1994evolution,fu2018spatial} bacterial populations, tissues and organs in multicellular organisms \citep{carroll2001chance,hedges2004molecular},
casts and social classes in colonial animals \citep{beshers2001models, smith2008genetic}, and guilds in ecological communities \citep{terborgh1986guilds,futuyma1988evolution,may1986ideas}, all fulfill specialized roles that are vital for the functioning of a larger whole.
Specialization also gives rise to metabolic interdependencies in microbial populations and can serve as a strong mechanism for community assembly \citep{zelezniak2015metabolic}. 

Evolution of specialization is typically studied in terms of fitness trade-offs or economic considerations. Specialization emerges if relatedness is high and if fitness returns accelerate \citep{michod2007evolution,michod2006life,willensdorfer2009evolution,tannenbaum2007does,rueffler2012evolution,taylor1992altruism,dcv2}. There are two classes of evolutionary forces moving a population from having one type of individual performing multiple functions $\text{--generalism--}$, towards one that has multiple types of individuals performing distinct functions $\text{--specialism--}$. The first is ``incompatible optimas'' \citep{solari2013general, sriswasdi2017generalist, goldsby2012task}: If a population must optimize two functions at once, but the phenotypes optimizing these are incompatible, then the population will split into two phenotypes. For example, the somatic and germ cells in volvox colonies are optimized for motility and reproduction. As a result, they have entirely different positioning \citep{solari2006hydrodynamics}, morphology \citep{kirk2001germ}, and protein expression \citep{kirk1983protein}. In multicellular cyanobacteria, cells differentiate into carbon-fixating cells and nitrogen-fixating heterocysts \citep{rossetti2010evolutionary}. E. coli can differentiate into transient non-growing cells and normally growing cells to hedge their bets across different environments \citep{lewis2007persister}. A traveling band of E. coli will exhibit a continuum of navigation styles, each specializing in processing different local conditions while still moving in unison \citep{fu2018spatial}.

A second type of evolutionary pressure originates from the economies of scale. Undertaking one process at high volume is more cost-effective than undertaking multiple processes at low volume. The morphological characteristics necessary to accomplish two distinct functions require two investments in overhead. Specialization is then favored if fitness returns are accelerated by further investment into a specific task \citep{west2015major, cooper2018division}. 



\begin{figure}
    \centering
    \includegraphics[width=0.49\textwidth]{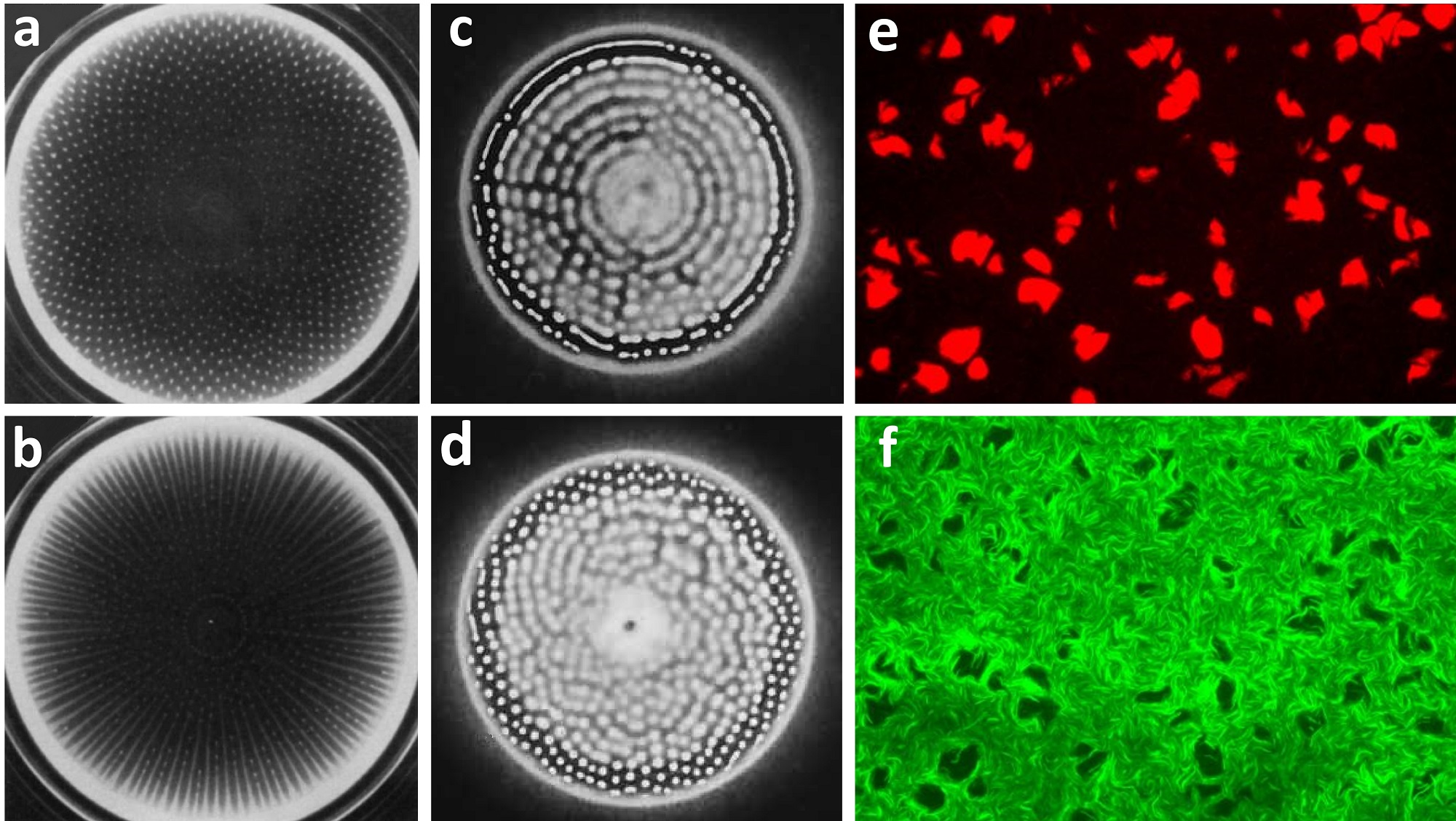}
    \caption{{\bf Pattern forming populations.} {\bf a,b} Examples of spatial patterning of bacteria in experiments performed by Budrene and Berg (1991) \citep{budrene1991complex}. E. coli formed spots ({\bf a}) and stripes ({\bf b}) in response to public goods they themselves excrete. {\bf c,d} Aggregation patterns observed in S. typhimurium in experiments by Blat and Eisenbach (1995) \citep{blat1995tar}. {\bf e,f} Spot ({\bf e}) and hole ({\bf f}) patterns observed in experiments with synthetic bacteria performed by Karig et al. (2018) \citep{karig2018stochastic}. Figures courtesy of authors.}
    \label{fig:patterns_ecoli}
\end{figure}

It is well known that spatial structure is key in the evolution of cooperation \citep{durrett1994importance,taylor1992altruism,lion2008self,wakano2009,uppal2018}. By forming fragmenting groups, multicellular organisms and social colonies can combat fixation of cheaters. Coexistence of cheaters and cooperators is also enhanced in spatially structured populations \citep{wilson2003coexistence}. Understanding how spatial structuring arises and competition within and across groups can shed light on how cooperation and resistence to cheaters arise \citep{lion2008self}. Here we will be interested in the role of spatial structuring in the evolution of specialization.

Existing computational models of evolution of specialization that consider spatial structure or finite group size, typically abstract away the underlying physics \citep{dcv2,cooper2018division,gavrilets2010rapid,willensdorfer2008organism,rueffler2012evolution, ispolatov2012division,menon2015public,gavrilets2010rapid, willensdorfer2008organism,oliveira2014evolutionary,schiessl2019individual}. While conceptually useful, such models reveal little about the interplay between evolutionary and mechanical forces during the formation and evolution of specialization. Real-life microbial exchanges are mediated almost entirely by viscoelastic secretions that diffuse and flow \citep{west2007social}. Extracellular enzymes digest food \citep{greig2004prisoner, bachmann2011high, pirhonen1993small}, surfactants aid motility \citep{kearns2010field,xavier2011molecular}, chelators scavenge metals \citep{griffin2004cooperation, guerinot1994microbial, ratledge2000iron,neilands, harrison,kummerli}, toxins fight competitors and antagonists \citep{mazzola1992contribution,moons2005quorum,moons2006role,an2006quorum,inglis2009spite}, virulence factors exploit a host \citep{zhu2002quorum, Allen2016, sandoz2007social, kohler}, and extracellular polymeric substances provide sheltering \citep{mah2001mechanisms,xue2012multiple,davies2003understanding}. Since cells must be within a certain distance to exchange such services, spatial aggregation is considered a prerequisite for multicellular specialization. Spatial effects matter \citep{durrett1994importance, wilson2003coexistence, fletcher2009simple, wakano2009,mcnally2017killing}, and multiple factors can couple together to influence the evolution of cooperation \citep{Dobay2014} and division of labor \citep{dragovs2018division} in unexpected ways.

In this study we find that mechanical factors such as diffusion constants, molecular decay rates and fluid flow patterns play a crucial role in shaping the interaction structure of an ecological community. We find, through first-principles computer simulations and matching analytical formulas, that microbes self-aggregate and form evolving clusters, whose size, shape and economical exchanges are sensitively dependent on the physical parameters defining the abiotic environment. Such structures have already been empirically observed in \emph{E. coli} \citep{budrene1991complex}, \emph{S. typhimurium} \citep{blat1995tar}, and \emph{B. subtilis} \citep{mendelson1998complex} (Fig. \ref{fig:patterns_ecoli}) and studied theoretically  \citep{tsimring1995aggregation,wakano2009,stump2018local,mcnally2017killing}. However, the interplay between evolutionary and mechanical forces within and between these structures and their role in the formation and evolution of community interactions remain unknown. 

Since many bacterial products leak outside the cell, members of the local community can exploit their neighbors and evolve to delete costly functions. The Black Queen Hypothesis suggests that loss of functionality occurs due to selfish mutations and can form the basis for mutualistic relationships \citep{morris2012black, sachs2012origins}. 
Thus, from evolutionary game theoretical considerations alone, one expects that specialists always eventually dominate a population of generalists. How then should we explain the persistence of generalists in nature, and even the coexistence of various combinations of generalists, specialists, and cheaters within one niche? 

To address this question we construct a mechanistic model that naturally gives rise to distinct microbial clusters. We then analyze the evolutionary transitions between generalized and specialized interactions within clusters for different fluid flow patterns, diffusion lengths, molecular decay constants and cell growth kinetics. Lastly, we study the competitive interactions across clusters.

In doing so we establish the physical factors that counteract game theoretical expectations, i.e. factors that allow generalists to resist specialization, and generalists and specialists to resist cheaters. We also establish physical factors that counteract competitive exclusion, i.e. allowing multiple community types to coexist within the same fluid niche. Lastly, we determine what physical properties make ``socially uninhabitable'' niches, where free-riders emerge, exploit and invariably destroy both generalist and specialist communities.

\section*{Methods}
\begin{figure*}
\centering
\includegraphics[width=\textwidth]{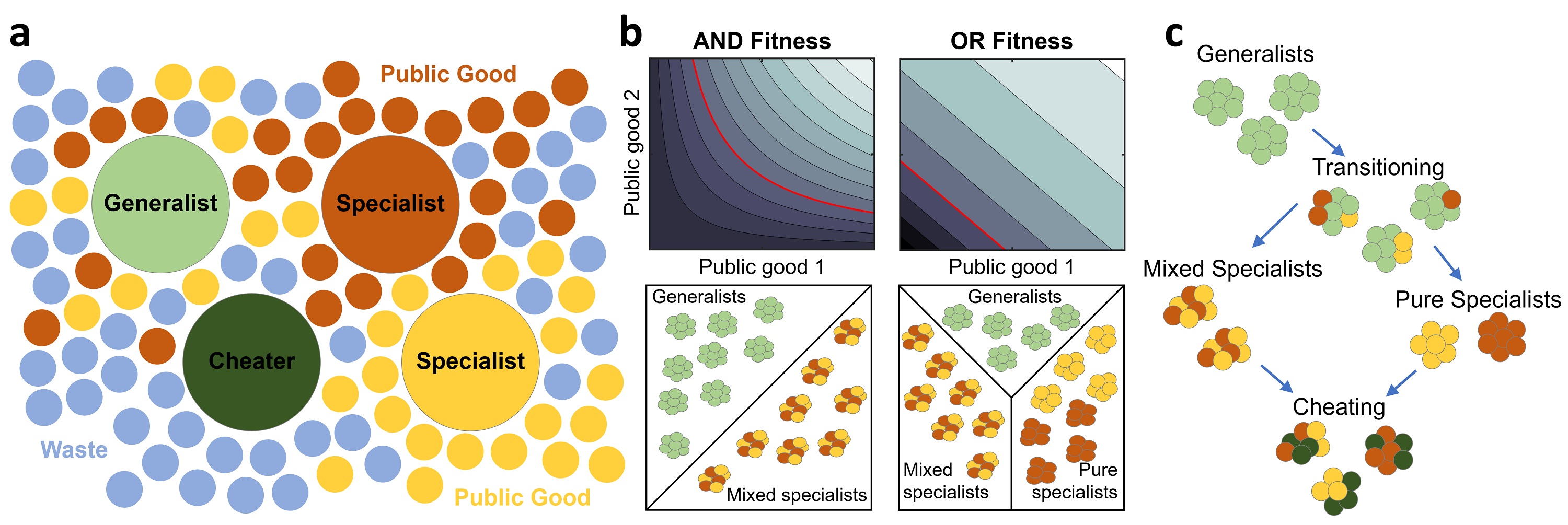}
\caption{{\bf Schematics and dynamics of the model.} {\bf a,} Microbes cooperate by secreting public goods into their environment. Generalists (large green circles) secrete two public goods (small yellow and red circles). Specialists (large red and yellow circles) secrete only one of the two public goods. Cheaters (dark green) secrete none of the public goods. All microbes secrete a metabolic waste (small blue circles). {\bf b,} Fitness contour plots and types of stable groups in each fitness variant. In the top row, we plot the fitness contours as a function of public good concentrations $c_1$ and $c_2$. In the bottom row, we show the types of stable groups in each fitness form. The red line represents the contour corresponding to zero fitness. In the AND case, the red line can never cross the $c_1, c_2$ axes, the fitness is negative when either chemical is not present. Therefore the only types of stable groups are generalists or mixed specialists, as shown in the bottom-left panel. In the OR case, the contours form straight lines. A decrease in one chemical is equally compensated by an increase in the other chemical. The zero contour always crosses the $c_1, c_2$ axes. It is therefore also possible to have pure specialist groups in the OR case, as shown in the bottom-right panel. {\bf c,} Evolutionary paths between group types. Groups typically move towards less secretion since cheaters have higher invasion fitness than specialists, who, being ``half-cheaters'', have higher invasion fitness than generalists.} 
\label{fig:model_schematic}
\end{figure*}

Any model aiming to describe evolution of functional specialization must include at least two functions, so that sub-populations can potentially specialize to perform one function each. In our model microbes can secrete two public goods and a waste/toxin. These molecules diffuse, flow, and decay (cf. Fig. \ref{fig:model_schematic}). 

The specific assumptions of our model, qualitatively stated, can be enumerated as follows: {\bf (1)} The system consists of microbes that can secrete two kinds of public goods. A public good refers to a secretion that promotes the growth of nearby microbes (including the producer). The producer also pays a metabolic cost for secreting the public good. {\bf (2)} Every microbe secretes a waste molecule that curbs the growth of those nearby. {\bf (3)} The secretions and bacteria obey the physical laws of fluid dynamics and diffusion. {\bf (4)} Whether a microbe secretes both, one, or none of the public goods is hereditary, except for mutations. However, every phenotype emits waste.


We study two models separately. {\bf (5)} In one, which we call AND, access to both kinds of goods is necessary. In the other, which we call OR, both goods contribute to fitness, but the lack of one can be compensated with the other.

Our work consists of discrete, stochastic agent based simulations and related continuous deterministic equations. In addition, to gain better analytical understanding, we construct a simple effective model that captures the essential outcomes of the simulations.

\subsection*{Continuous Deterministic Equations}
We construct equations governing the number density of four phenotypes $n_0(x,t)$, $n_1(x,t)$, $n_2(x,t)$, $n_3(x,t)$ two chemical secretions that are public goods $c_1(x,t), c_2(x,t)$, and a waste compound $c_3(x,t)$, as a function of space $x$ and time $t$. $n_3(x,t)$ is the number density of microbes that secrete both kinds of public goods, to which we refer as ``generalists''. The microbes that secrete only public good one or two, are denoted by $n_1(x,t)$ and $n_2(x,t)$, to which we refer as ``specialists''. Those that secrete no public goods are denoted by $n_0(x,t)$, to which we refer as ``cheaters''. 
\begin{align}
&\dot{n}_i \!=\!  \left(d_b \nabla^2 \!-\! \mathbf{v}(x,t) \!\cdot\! \boldsymbol{\nabla} \!+\! f_i (\mathbf{c}) \right) n_i \!+\! \sum_{j =0}^3 M_{ij} n_j  
\label{eq:microbes} \\
&\dot{c}_\alpha \!=\! \left( d_\alpha \nabla^2 \!-\! \mathbf{v}(x,t) \!\cdot\! \boldsymbol{\nabla} \!-\! \lambda_\alpha \right) c_\alpha \!+\! \sum_{i=0}^3 S_{i \alpha} n_i 
\label{eq:chemicals}
\end{align}
Here indices $i,j = 0,1,2,3$ label phenotypes, whereas the index $\alpha = 1,2,3$  labels chemicals, i.e. the two public goods and waste. Thus, Eqn.(\ref{eq:microbes}) and (\ref{eq:chemicals}) comprise 7 coupled spatiotemporal equations.

In both equations, the first two terms describe diffusion and advection. The flow field $\mathbf{v}(x,t)$ is a vector valued function of space and time, and includes all information pertaining the flow patterns in the environment. In general, it is obtained by solving separate fluid dynamics equations. Mutations and secretions are governed by two matrices,
\begin{align*}
M =  \mu
\begin{bmatrix}
-2   & 1   &  1  &  0 \\
1    & -2  &  0  &  1 \\
1    & 0   & -2 &  1 \\
0    &  1  &  1  & -2 
\end{bmatrix}, \qquad
S =  
\begin{bmatrix}
0   &  0   &  s_w \\
s_1 &  0   &  s_w \\
0   &  s_2 &  s_w \\
s_1 &  s_2 &  s_w 
\end{bmatrix} .
\end{align*}
The secretion rate of chemical $\alpha$ by phenotype $i$ is given by the matrix element $S_{i \alpha}$, and its decay rate by $\lambda_\alpha$. The mutation rate from phenotype $j$ to $i$ is given by $M_{ij}$. The diagonal elements $M_{ii}$ indicate the rate at which $i$ mutates to become something else.


Note that in our model, the secretion of public goods is binary, i.e. a good is either secreted or not. Mutations toggle on and off with probability $\mu$ whether an individual secretes either public good. A mutation can cause a generalist to become a specialist, but two mutations, one for each secretion function, are required for a generalist to become a cheater. Same with back mutations. 

The fitness function $f_i (\mathbf{c})$ determines the growth rate of phenotype $i$. We consider two cases separately: when both public goods are necessary for growth (AND) and when the public goods can substitute one-to-one for one other (OR). 
\begin{align}
&f^{(\text{AND})}_i \!=\! a_{12} \frac{c_1 c_2}{c_1 c_2 + k_{12}} \!-\! a_w \frac{c_3}{c_3 + k_w} \!-\!  \sum_{\alpha = 1}^2 \beta_\alpha S_{i \alpha} 
\label{eq:fAND} \\[2mm]
&f^{(\text{OR})}_i  \!=\! a_{12} \frac{(c_1 + c_2)}{(c_1 + c_2) + k_{12}}  \!-\! a_w \frac{c_3}{c_3 + k_w} \!-\! \sum_{\alpha = 1}^2 \beta_\alpha S_{i \alpha} 
\label{eq:fOR} 
\end{align}
As we see, in both cases, growth rate increases with the local concentration of public goods, $c_1, c_2$ and decreases with the concentration of waste, $c_3$. $\beta_\alpha$ is the cost of secreting public good $\alpha$, so that growth of phenotype $i$ is curbed by an amount proportional to its public good secretion. Waste is produced without any cost.
\begin{table*}
\caption{\label{tab:parameters} Summary of system parameters.}
\hspace{-0.1in}\begin{tabular}{l l l l}
\hline
{\bf } & {\bf Quantity} 					& {\bf Values for OR}  & {\bf Values for AND}  \\ 
\hline \vspace{-3mm} \\ 
$d_b$ 	 		& Microbial diffusion 		& $0.4  \times 10^{-4} \ \mathrm{cm}^2 \ \mathrm{s}^{-1}$ & $1 \!\times\!10^{-6} \ \mathrm{cm}^2 \ \mathrm{s}^{-1}$ \\
$d_1$	 	 	& Good 1 diffusion 	& $ 5 \text{ and } 25  \times 10^{-6} \ \mathrm{cm}^2 \ \mathrm{s}^{-1}$ & $ 5 \text{ and } 20 \!\times\!10^{-6} \ \mathrm{cm}^2 \ \mathrm{s}^{-1}$\\
$d_2$	 		& Good 2 diffusion 			& $5 \text{ and } 25  \times 10^{-6} \ \mathrm{cm}^2 \ \mathrm{s}^{-1}$ & $5 \text{ and } 20\!\times\!10^{-6} \ \mathrm{cm}^2 \ \mathrm{s}^{-1}$\\
$d_w$	 		& Waste diffusion 			& $10 \text{ to } 80  \times 10^{-6} \ \mathrm{cm}^2 \ \mathrm{s}^{-1}$ &$10 \text{ to } 80\!\times\!10^{-6} \ \mathrm{cm}^2 \ \mathrm{s}^{-1}$\\
$\lambda_1$		& Good 1 decay 		& $5.0\! \times\!10^{-3} \ \mathrm{s}^{-1}$ & $5.0\!\times\!10^{-3} \ \mathrm{s}^{-1}$\\
$\lambda_2$		& Good 2 decay 		& $5.0\!\times\!10^{-3} \ \mathrm{s}^{-1}$ & $5.0\!\times\!10^{-3} \ \mathrm{s}^{-1}$ \\
$\lambda_w$		& Waste decay 				& $1.5\!\times\!10^{-3} \ \mathrm{s}^{-1}$ & $1.5\!\times\!10^{-3} \ \mathrm{s}^{-1}$\\
$k_{12}$ 		& Goods saturation			& $0.01$ & $3\!\times\!10^{-5}$ \\
$k_w$ 		 	& Waste saturation 					& $0.1$ & $0.1$ \\
$s_1$ 		 	& Good 1 secretion rate		& $5.0\!\times\!10^{-3} \ \mathrm{s}^{-1}$ & $0.01 \ \mathrm{s}^{-1}$\\
$s_2$ 		 	& Good 2 secretion rate		& $5.0\!\times\!10^{-3} \ \mathrm{s}^{-1}$ & $0.01 \ \mathrm{s}^{-1}$\\
$s_w$ 		 	& Waste secretion rate 				& $0.01 \ \mathrm{s}^{-1}$ & $0.09 \ \mathrm{s}^{-1}$ \\
$a_{12}$ 		& Benefit from goods 			& $62.5 \text{ to } 80 \!\times\!10^{-3} \ \mathrm{s}^{-1}$ & $40 \text{ to } 75 \!\times\!10^{-3} \ \mathrm{s}^{-1}$ \\
$a_w$  		& Harm from waste 			& $8.0 \!\times\!10^{-3} \ \mathrm{s}^{-1}$ & $10.5 \!\times\!10^{-3} \ \mathrm{s}^{-1}$ \\
$\beta_1$ 			& Cost of good 1				& $0.01$ to $0.26$ & $0.01$ to $0.15$\\
$\beta_2$ 			& Cost of good 2				& $0.01$ to $0.26$ & $0.01$ to $0.15$\\
$\mu$ 			& Mutation rate 					& $5.0\!\times\!10^{-8} \ \mathrm{s}^{-1}$ & $2.0\!\times\!10^{-7} \ \mathrm{s}^{-1}$\\

\end{tabular}\vspace{-0.2in}
\end{table*}

Note that with increasing concentration of goods, microbes receive diminishing returns. Similarly, with larger waste, death rate approaches a maximum value. These functional forms are well understood, experimentally verified \citep{Monod1949}, and commonly used in population dynamics models \citep{allen2018bacterial}. $a$'s and $k$'s are constants defining the initial slope and saturation values of growth and death (see Table \ref{tab:parameters}).

\subsection*{Discrete Stochastic Simulations}
Our analytical conclusions (cf. Supplementary Section \ref{app:turing}) have been guided and supplemented by agent based stochastic simulations in two dimensions. Videos of these simulations are provided in Supplementary Videos. Our simulation algorithm is as follows: at each time interval, $\Delta t$, the microbes (1) diffuse by a random walk of step size $\delta = \sqrt{4 d_b \Delta t} + \vec{v} \Delta t$ derived from the diffusion constant plus a bias dependent on the flow velocity. (2) Microbes secrete chemicals locally onto a discrete grid that then diffuse using a finite difference scheme. (3) Microbes reproduce or die with a probability dependent on their local fitness and time step, given by $f(\mathbf{c}) \Delta t$. If $f \Delta t$ is negative, the microbes die with probability 1, if $f \Delta t$ is between 0 and 1 they reproduce an identical offspring with probability $f \Delta t$. Upon reproduction, offspring are placed at the same location as their parent. (4) Random mutations may alter the secretion rate of either public good --and thus the reproduction rate-- of the microbes. Mutations occur on each secretion function with probability $\mu$ and turn the secretion of the public good on or off. The secretion rate is assumed to be heritable, and constant in time. Numerical simulations for figures were performed by implementing the model described above using the Matlab programming language and simulated using Matlab (Mathworks, Inc.). The source code for discrete simulations is provided as a supplemental file. Additional details of model implementation are discussed in Supplementary Section \ref{app:para}.

A summary of the system parameters is given in Table \ref{tab:parameters}, along with typical ranges for their values used in the simulations. Parameter values as well as the simulation domain (the physical region being simulated) are also given in figure captions. The relevant ratios of parameters are consistent with those observed experimentally \citep{Kim1996,Ma2005}. Note also that the choice of parameters will be restricted to ensure a finite stable solution is possible. For example, we enforce the quantity $a_{12} - a_{w} - \beta_1 s_{1} - \beta_2 s_{2} < 0$. This is because, if this quantity were positive, then a dense population where the Hill terms in the fitness functions are saturated, will continue to have a positive fitness and grow indefinitely. In the case where secretion rate and/or production costs are low, the waste term is crucial to ensure a finite carrying capacity. We therefore choose $a_w \geq a_{12}$. Other constraints on existence and stability are derived in our Turing analysis (see Supplementary Section \ref{app:turing}). Further discussion on parameter selection and sensitivity is also given in Supplementary Section \ref{app:para}.

\subsection*{Simple Effective Model}
To gain better analytical understanding, we set to reproduce the outcomes our complex model with a much simpler effective model, which we describe in Supplementary Section \ref{app:effectiveGroup}. Our effective model is based on the observation that microbes aggregate into self-reproducing cooperative groups. Different group types, rather than individual microbes, constitute the basic building blocks of our effective model; and the fragmentation rates of these group types constitute the basic parameters of the model. These parameters are ``measured'' from our complex simulations and depend on the physical properties of the system (see Supplementary Figures \ref{fig:effective_and},\ref{fig:effective_or}). The results of our effective model are compared to simulation results in Fig. \ref{fig:betaShear}.

\section*{Results}
\subsection*{Cooperative groups as Turing patterns}
\begin{figure*}
\centering
\includegraphics[width=\textwidth]{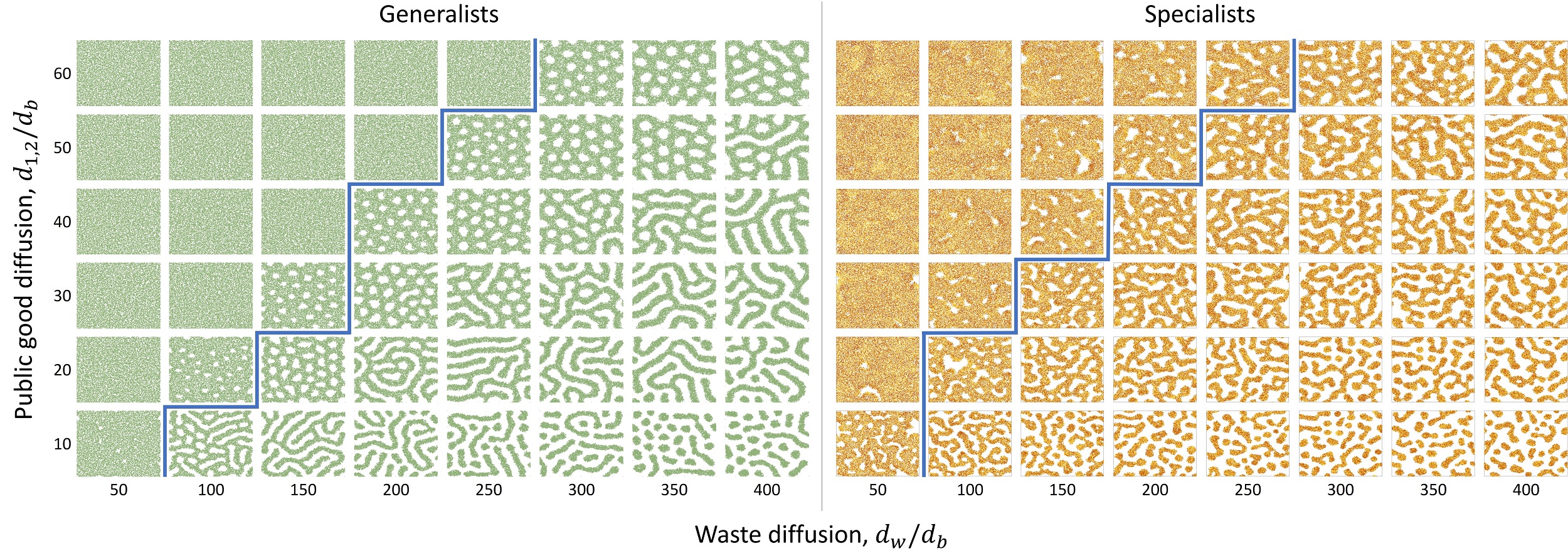}
\caption{{\bf Pattern formation in bacteria.}  In both AND and OR fitness types, microbes can form clusters. We set both the mutation rate and fluid flow rate to zero to study pattern formation in generalist and specialist microbes without additional physical and evolutionary complications. While this figure only shows generalists and specialists subject to AND-type fitness, we observe qualitatively identical patterns for the OR-type fitness. The thick blue line is the result of our analytical calculation indicating the regime where patterns emerge, which agrees with computer simulations (see Supplementary Section \ref{app:turing}). The population can be homogeneous or form stripes or spots. These patterns can also grow and fragment, forming new colonies. The diffusion constants are normalized by the bacterial diffusion constant, $d_b = 1 \times 10^{-6} \ \mathrm{cm}^{2} \mathrm{s}^{-1}$. Secretion constants used are $s_1 = s_2 = 0.015 \ \mathrm{s}^{-1}$ and $s_w = 0.005 \ \mathrm{s}^{-1}$, public good cost is $\beta_1 = \beta_2 = 0.1$. The rest of the parameters are as given in Table \ref{tab:parameters}. }
\label{fig:turing}
\end{figure*}

Through numerical simulations and analytical formulas, we see that the system gives rise to spatially segregated cooperating groups in a certain parameter range, as shown in Fig. \ref{fig:turing}. Spots or stripes in reaction diffusion systems are known as Turing patterns, which form whenever an inhibiting agent diffuses faster than an activating agent. In our model the inhibiting and activating agents are the waste and the public goods. 

In general, the structure and size of these cooperating groups will vary with physical parameters. We show in Fig. \ref{fig:turing} how the Turing pattern forming region varies with diffusion constants, in the absence of mutations or flow. Our analytical result, derived in Supplementary Section \ref{app:turing}, shown by the thick blue lines, delineate the parameter space into pattern forming and non-pattern forming regions. While simulations agree well with analytical results, we see some patterns slightly beyond the theoretical region. This is due to the stochastic nature of the simulations which is known to widen the pattern forming region \citep{butler2009robust, biancalani2010stochastic}. 

In our simulations, we observe that cooperative groups of microbes, i.e. spots and stripes, grow and fragment, thereby giving rise to new structures of the same type. The spatial structure of these patterns differ between generalists and specialists, and therefore have a strong effect on the evolutionary trajectory of the system.

\subsection*{Effects of secretion cost on specialization}
We next determine the role of secretion cost $\beta_\alpha$ on group structure and hence specialization, in the absence of flow. To see the effect of trade-offs on specialization, we varied the cost of public good secretion and determined when specialization occurs in both AND and OR fitness forms. To simplify our analysis, we set $s_1 = s_2 = s$. In order for both types of specialists to then coexist, we also set $\beta_1 = \beta_2 = \beta$. Therefore, generalists pay an overall cost of $2 \beta$, specialists pay $\beta$, and cheaters pay no cost. As such, a specialist mutant will invade a generalist group, and a cheater mutant will invade a specialist group. In the absence of spatial structure and flow, the entire population will be dominated by cheaters and will go extinct.

\begin{figure*}
\centering
\hspace{-0.1in}\includegraphics[width=0.95\textwidth]{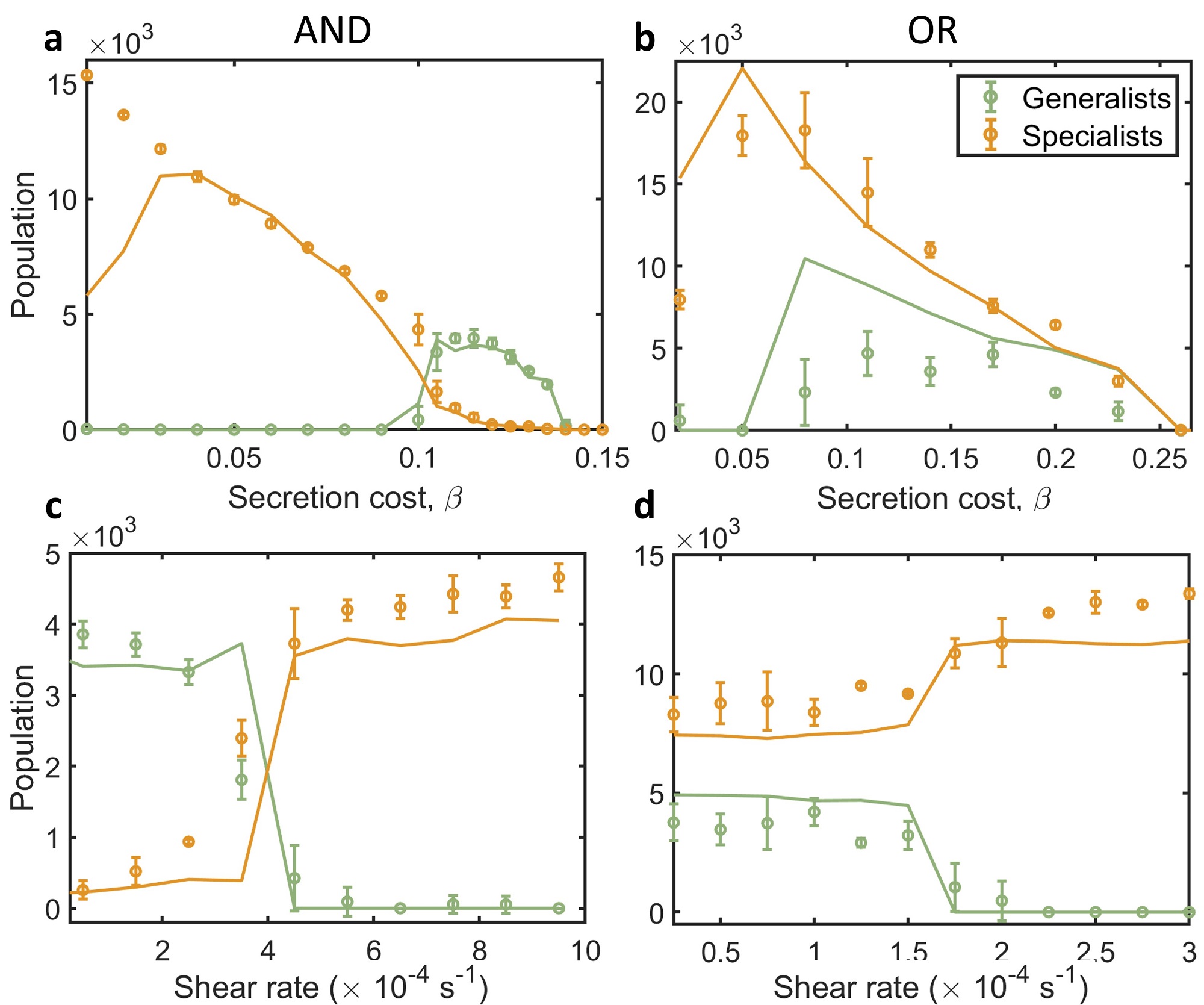}
\caption{{\bf Effects of cooperation cost and fluid shear on specialization.} Points with error bars represent numerical results averaged over 5 runs at simulation time $T=2 \times 10^{6} \ \mathrm{s}$ in a domain of size 20 $\mathrm{mm}$ $\times$ 20 $\mathrm{mm}$. Error bars correspond to one standard deviation from the mean. Solid lines are from our effective theoretical model given in Supplementary Section \ref{app:effectiveGroup}. {\bf a,b,} Effects of secretion costs in the absence of flow. In each fitness variant, we see that specialization is more abundant for low costs. At lower costs, generalist groups are ``too fit'' and form large aggregate structures that are more susceptible to mutations. At higher costs, {\bf a,} we see generalists out-compete specialists, in the AND case (see Supplementary Video 1) and {\bf b,} coexist with specialists in the OR case (Supplementary Video 5). At low costs, specialists again dominate. {\bf c,d,} Effects of fluid shear. A shearing flow causes groups to fragment quicker and stripes to elongate and grow larger. {\bf c,} In the AND case, with secretion cost $\beta = 0.12$, we observe that shear transitions the system from a coexisting state to a specialist state (see Supplementary Video 2). Here, shear causes generalists groups to elongate and become more susceptible to mutations. {\bf d,} In the OR case, at cost $\beta = 0.17$, shear again causes a transition from a coexisting state to a pure specialist state. We therefore see in both cases that flow shear will promote specialization. Parameter values for diffusion are $d_1=$ $d_2=5 \times 10^{-6} \ \mathrm{cm}^2 \ \mathrm{s}^{-1},$ $d_w = 15 \times 10^{-6} \ \mathrm{cm}^2 \ \mathrm{s}^{-1}$. For public good benefit, in the AND case, $a_{12} = 6.5\times 10^{-3} \ \mathrm{s}^{-1}$, and in the OR case, $a_{12} = 7.5 \times 10^{-3} \ \mathrm{s}^{-1}$. The rest of the parameter values are given in Table \ref{tab:parameters}.}
\label{fig:betaShear}
\end{figure*}

What can we say about the competition between different group types (as opposed to between different strains within a group)? Since with all else equal, increasing costs harm generalists twice as much as specialists, one might expect that increasing the cost of the goods would favor the specialists over generalists.
Counterintuitively, we find the opposite. Specialist groups indeed grow faster and form larger, expansive, and denser groups, which however are at once taken over by cheaters. In contrast, generalists form smaller, sparser, weaker groups that fragment more often, which limits the spread of mutants (see Supplementary Figures \ref{fig:effective_and},\ref{fig:effective_or}). Therefore, at higher cost $\beta$, the ``weak'' generalists are able to coexist and even dominate ``strong'' specialists (Fig. \ref{fig:betaShear}a,b). 

In general, a large uniform population is more susceptible to invading mutants. In contrast, when the population is organized as fragmenting patches, the community structure will prevail as long as the fragmentation rate is larger than the invasive mutation rate. Thus, the type, size, growth and fragmentation of the groups ultimately dictates whether generalism, specialism, or a coexistence of group types are evolutionarily stable.


 \subsection*{Effect of flow patterns on specialization}
Fluid dyanmical forces can strongly influence the eco-evolutionary dynamics of a microbial population. For example, fluid flows can shape the competition and matrix secretion in biofilms \citep{nadell2017flow}. A shearing fluid flow has also been shown to modify social behavior by enhancing the group size and fragmentation rate \citep{uppal2018}. We therefore expect that the flow patterns will affect the mode of cooperation (specialist vs. generalist) and the physical structure of groups.

For constant shear we used a planar Couette flow, with velocity profile and shear rate given as,
\begin{align*}
    \mathbf{v} = v_{\text{max}} \frac{y}{H} \hat{\mathbf{x}}, \quad \left| \frac{\mathrm{d} v}{\mathrm{d} y} \right| = \frac{v_{\text{max}}}{H},
\end{align*}
where $v_{\text{max}}$ is the maximum flow rate and $H$ is the height of the domain. Flow is along the $\hat{x}$ direction and is zero in the center $y = 0$, and maximal at the boundaries $y = \pm H$. We used periodic boundary conditions along the left and right walls ($\hat{\mathbf{x}}$ direction), and Neumann boundary conditions for the top and bottom surfaces ($\hat{\mathbf{y}}$ direction).

The effect of shear is in general non-trivial and will depend on the group structure observed. We find that a shearing flow increases group fragmentation rate of microbes organized in distinct circular spots, whereas it simply enlarges groups when they are organized in an elongated, stripe-like fashion.

In Fig. \ref{fig:betaShear}c,d we show the effect of shear at intermediate costs, where its effect is strongest. We found in both cases that larger shear helps specialists by enhancing their fragmentation rate and enlarging generalist groups (Supplementary Figures \ref{fig:effective_and},\ref{fig:effective_or}), since larger generalists groups generate more mutations, and since faster fragmenting specialist groups are better able to resist takeover by cheaters. Here, fluid shear transitions the system from a generalist or coexisting state to a specialist state (Supplementary Video 2). Thus, fluid shear promotes specialization. 

Since advective flow is something that one can tune in an experimental or industrial setting, it is exciting to think of possibilities where flow is used to control the social evolution of a microbial community. Furthermore, since shear is in general spatially dependent, we can use different velocity profiles to localize this control to different regions. 


\subsection*{Effect of public good benefit, cooperation cost, and competition on evolution of specialization}
\begin{figure*}
\centering
\includegraphics[width=0.92\textwidth]{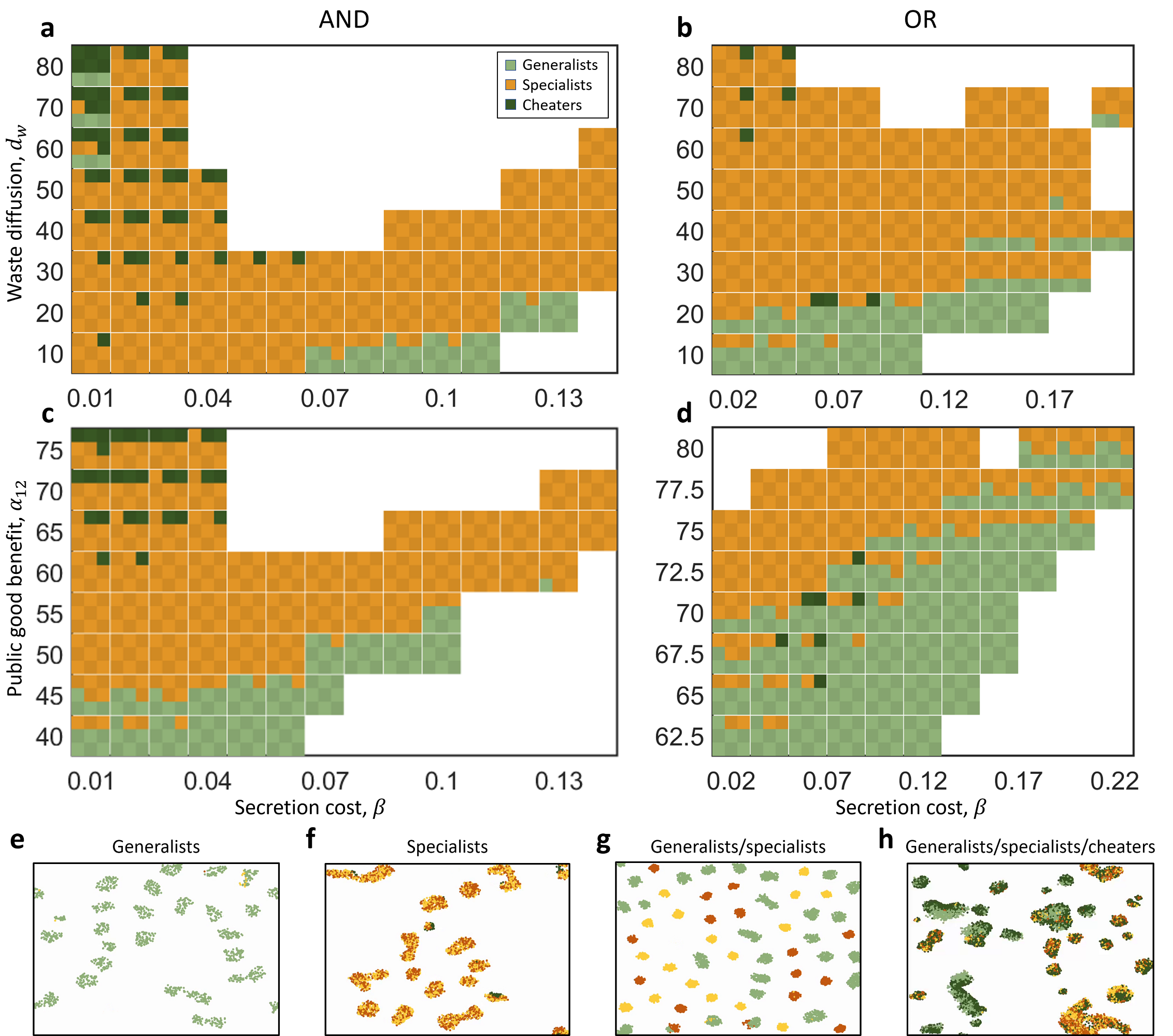}
\caption{{\bf Effects of waste diffusion, public good benefit, and cooperation cost on specialization.}  {\bf a,b,} Population composition for varying secretion cost $\beta$ and waste diffusion $d_w$ for AND ({\bf a}) and OR fitness types ({\bf b}). Each square is filled proportionally to the population composition of generalists, specialists, and cheaters. When waste diffusion is larger than the public good diffusion, the population will form spatial structures. Under certain conditions, we see that generalists, specialists, and cheaters coexist. At low costs, specialization and cheating are more abundant. At medium costs, cheaters spread faster than groups fragment, leading to extinction, shown as empty regions. At higher costs, specialists form smaller groups that fragment quicker than cheaters spread and are stable at steady state. Specialists do better overall when waste diffusion is large, since they can then form denser groups without over-polluting themselves. {\bf c,d,} Population compositions for varying secretion cost $\beta$ and public good benefit $a_{12}$ for AND ({\bf c}) and OR fitness types ({\bf d}). Higher public good benefit, $a_{12}$, also helps specialization, since secreting fewer public goods still gives a large benefit. Interestingly, higher benefit also leads to more extinct states, since cheaters can take over quicker. {\bf e-h,} Simulation snapshots for various possible stable populations. Under certain conditions we can see {\bf e,} stable generalists, shown here for AND fitness (Supplementary Video 1). Parameters here are $\beta = 0.12, d_1 = d_2 = 5 \times 10^{-6} \ \mathrm{cm}^2 \ \mathrm{s}^{-1}, d_w = 15 \times 10^{-6} \ \mathrm{cm}^2 \ \mathrm{s}^{-1}, a_{12} = 6.5 \times 10^{-3} \ \mathrm{s}^{-1}$. {\bf f,} Stable specialists in AND fitness (Supplementary Video 3). Parameters here are $d_1 =$ $d_2 = 5 \times 10^{-6} \ \mathrm{cm}^2 \ \mathrm{s}^{-1}$, $d_w = 15 \times 10^{-6} \ \mathrm{cm}^2 \ \mathrm{s}^{-1}$, $a_{12} = 6.5 \times 10^{-3} \ \mathrm{s}^{-1}$, $\beta = 0.08$. {\bf g,} Generalists coexisting with specialists in OR fitness (Supplementary Video 5). Parameter values $d_1 =$ $d_2 = 25 \times 10^{-6} \ \mathrm{cm}^2 \ \mathrm{s}^{-1}$, $d_w = 40 \times 10^{-6} \ \mathrm{cm}^2 \ \mathrm{s}^{-1}$, $a_{12} = 6.5 \times 10^{-3} \ \mathrm{s}^{-1}$, $\beta = 0.14, \mu = 5 \times 10^{-8} \ \mathrm{s}^{-1}$ {\bf h,} Coexistence of generalists, specialists, and cheaters in AND fintess (Supplementary Video 6). Parameter values $d_1 = d_2 = 20  \times 10^{-6} \ \mathrm{cm}^2 \ \mathrm{s}^{-1}$, $d_w = 60  \times 10^{-6} \ \mathrm{cm}^2 \ \mathrm{s}^{-1}$, $a_{12} = 7.5 \times 10^{-3} \ \mathrm{s}^{-1}$, $a_w = 105 \times 10^{-3} \ \mathrm{s}^{-1}$, $\beta = 0.02, \mu = 2 \times 10^{-7}  \ \mathrm{s}^{-1}$. Population values were obtained by taking a time average over 1 run for each parameter value, over time steps $T=1 \times 10^{6} \ \mathrm{s}$ to $T=2 \times 10^{6} \ \mathrm{s}$. Public good diffusion for the AND case ({\bf a,c}) is $d_{12} = 20 \times 10^{-6} \ \mathrm{cm}^2 \ \mathrm{s}^{-1}$, for the OR case ({\bf b,d}) $d_{12} = 25 \times 10^{-6} \ \mathrm{cm}^2 \ \mathrm{s}^{-1}$. Flow rate is set to 0 and other parameters are as given in Table \ref{tab:parameters}.}
\label{fig:pieGraphs}
\end{figure*}

We next study how varying public good benefit, production cost, and waste diffusion affect the stability of different community structures (Fig. \ref{fig:pieGraphs}). We find that higher waste diffusion and public good benefit helps specialists and higher secretion cost favors generalists. Fig. \ref{fig:pieGraphs} also shows what conditions leads to coexistence of different group types.

If waste diffusion is large, self-competition is lower, and specialists can form denser groups without over-polluting themselves (top regions in Fig. \ref{fig:pieGraphs}a,b). They can then better utilize public goods secreted by their neighbors. If the public good benefit, $a_{12}$, is large, specialists also do better since secreting fewer public goods still gives a large benefit (top regions in Fig. \ref{fig:pieGraphs}c,d, see also Supplementary Video 3 for AND fitness variant and Supplementary Video 4 for OR fitness). 

As we have already seen, specialization emerges when trade-offs are small, i.e. at smaller $\beta$. At higher $\beta$, generalists are able to coexist with specialists (see Fig. \ref{fig:pieGraphs}g and Supplementary Video 5 for OR fitness) and constitute the majority of the population (Fig. \ref{fig:pieGraphs}e and Supplementary Video 1). 

We also see that cheaters can persist stably with the population when their invasion fitness is lower than the growth rate of producers. This occurs in regions where producers do not form groups but grow either as stripes or homogeneously in space, which happens when public good benefit is large and when secretion costs are low. In this case, cheaters ``chase after'' producers, which grow into free space (Fig. \ref{fig:pieGraphs}h and Supplementary Video 6). 
High waste diffusion also helps cheaters, since they are able to chase producers without over-polluting themselves or their hosts (top-left regions in Fig. \ref{fig:pieGraphs}a,b). When their invasion fitness is about equal to the producer growth rate, cheaters take over fully, driving the population to extinction (top-center regions in Fig. \ref{fig:pieGraphs}a,b). When the population aggregates into groups, cheater growth is limited to the group. Cooperation then prevails if groups reproduce faster than cheaters emerge. This happens when secretion costs are large. Remarkably, higher secretion costs can therefore stabilize specialist populations against cheater invasion (top-right regions in Fig. \ref{fig:pieGraphs}a,b), since higher costs yield smaller groups which generate fewer mutations.

We see two regions of extinction: when public good benefit and waste diffusion are large, at medium costs (top-center regions in Fig. \ref{fig:pieGraphs}a-d); and when public good benefit and waste diffusion are low, at high costs (bottom-right regions in Fig. \ref{fig:pieGraphs}a-d). The first case is due to cheaters taking over groups, leading to the tragedy of the commons. Interestingly, this occurs more with higher public good benefit. The population of producers becomes ``too fit'' and more vulnerable to cheating mutations. For the second case, since costs are high and benefits are low, microbes need to form dense groups to utilize enough goods to be stable. However, due to the low waste diffusion, these groups over-pollute themselves and are no longer stable.

We see similar trends for both the AND (Fig. \ref{fig:pieGraphs}a,c) and OR cases (Fig. \ref{fig:pieGraphs}b,d). The main distinction between the two being, for the OR case, we predominately see pure specialist groups and only have mixed specialists in the AND case. We do not see many mixed specialists in the OR case since mutations take over generalists groups quicker and stabilize as pure groups, whereas in the AND case, pure groups would die out unless the complementary specialist also evolves in the same group. The AND structure is therefore essential to have true division of labor, where each type of specialist exists equally in the group.
\begin{figure*}
\centering 
\hspace{-0.15in}\includegraphics[width=\textwidth]{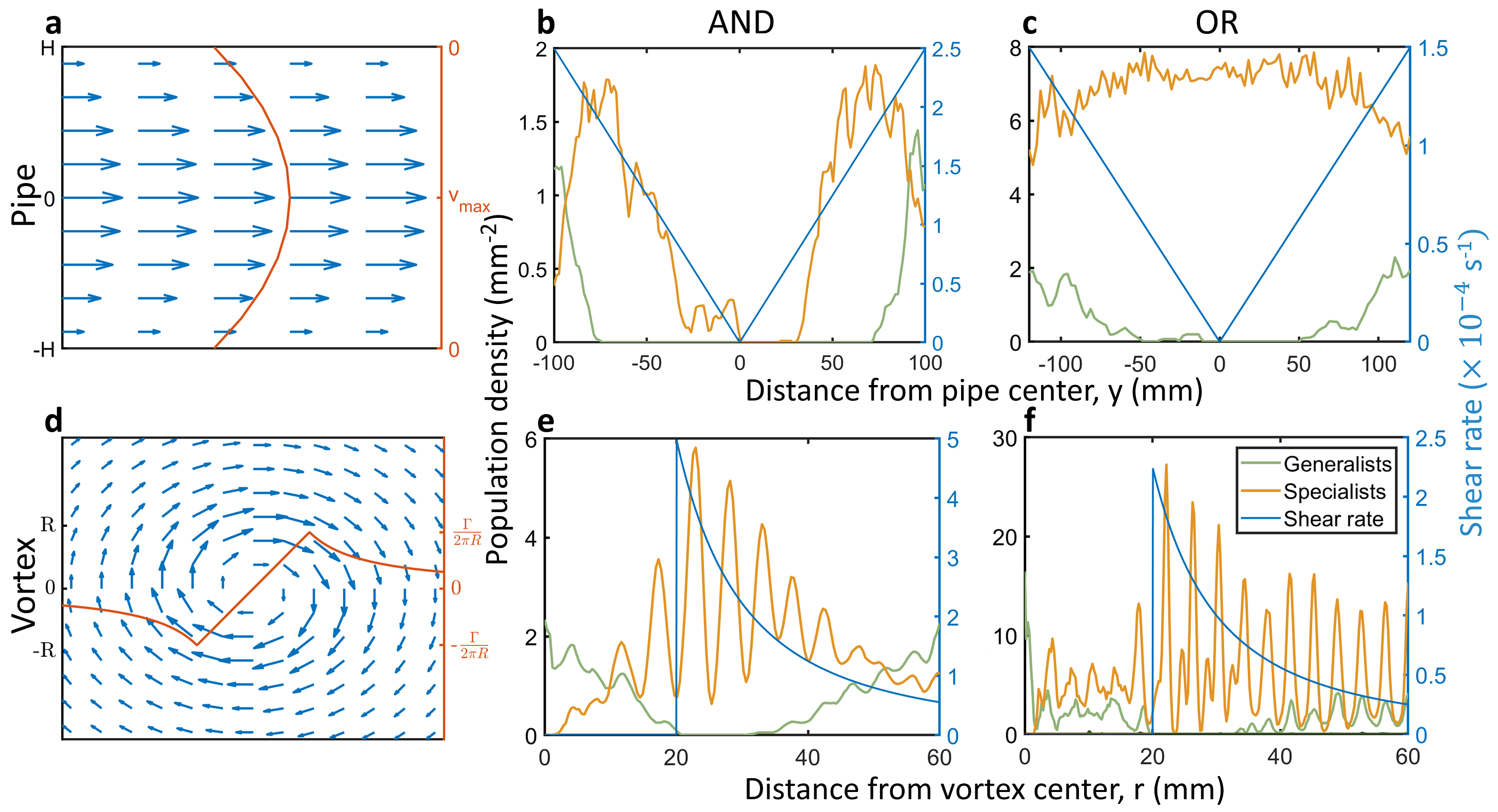}
\caption{{\bf Coexistence of specialists and generalists in pipes and vortices.} The local shear profile dictates which interaction structure is stable. When shear is spatially varying, we can get coexistence of generalists and specialists. {\bf a} Schematic of Hagen-Poiseuille flow in a 2-dimensional pipe. {\bf b} In the AND case in a pipe flow, we observe generalists residing at the boundaries followed by specialists towards the middle. In the center where shear is lowest, cheaters quickly spread and consume groups, leading to a local tragedy of the commons. The population then goes extinct in the center region (cf. Supplementary Video 7). Flow parameters are $H = 100 \ \mathrm{mm}$, $v_{\text{max}} = 125 \ \mathrm{mm} \ \mathrm{s}^{-1}$ and mutation rate $\mu = 5 \times 10^{-7} \ \mathrm{s}^{-1}$. {\bf c,} For the OR case, in a pipe, generalists and specialists coexist at the boundary while specialists dominate the center. Flow parameters for the OR case are $H=120 \ \mathrm{mm}$,  $v_{\text{max}} = 90 \ \mathrm{mm} \ \mathrm{s}^{-1}$ and mutation rate $\mu = 5 \times 10^{-8} \ \mathrm{s}^{-1}$. {\bf d,} Schematic of flow profile in Rankine vortex. {\bf e,} In a Rankine vortex flow, in the AND case we see generalists where shear is lowest, and specialists residing in an annulus where shear is at its maximum (cf. Supplementary Video 8). Flow parameters for the vortex in the AND case are $R = 20 \ \mathrm{mm}$, $\Gamma = 4000 \pi \ \mathrm{mm}^{2} \ \mathrm{s}^{-1}$, and mutation rate $\mu = 3 \times 10^{-7} \ \mathrm{s}^{-1}$. {\bf f,} In the OR case we see similar results, with coexistence of groups at low shear regions and an annular region composed of specialists. Flow parameters for the OR case are $R = 20 \ \mathrm{mm}$ and $\Gamma = 1800\pi \ \mathrm{mm}^{2} \ \mathrm{s}^{-1}$ and mutation rate $\mu = 5 \times 10^{-8} \ \mathrm{s}^{-1}$. The total simulation domain for vortices was $60 \ \mathrm{mm} \times 60 \ \mathrm{mm}$. Secretion costs used are $\beta = 0.12$ for the AND fitness, and $\beta = 0.17$ for the OR fitness. Diffusion parameters used were $d_1 = d_2 = 5\times 10^{-6} \ \mathrm{cm}^2 \ \mathrm{s}^{-1}, d_w = 15 \times 10^{-6} \ \mathrm{cm}^2 \ \mathrm{s}^{-1}$. Public good benefit in the AND case was $a_{12} = 6.5 \times 10^{-3} \ \mathrm{s}^{-1}$ and for the OR case, $a_{12} = 7.5 \times 10^{-3} \ \mathrm{s}^{-1}$. The rest of the parameters are as given in Table \ref{tab:parameters}. Population densities were obtained from averaging 5 runs at simulation time of $T=2 \times 10^6 \mathrm{s}$. }
\label{fig:pipe-vortex}
\end{figure*}

\subsection*{Localization of specialization and coexistence in axial and circular flows}
We next study the evolution of specialization in axial (Hagen-Poiseuille, Fig. \ref{fig:pipe-vortex}a) and circular (Rankine vortex, Fig. \ref{fig:pipe-vortex}d) flows. Again, we set the cost parameter to a value where shear makes the biggest difference. As with the case with constant shear (Fig. \ref{fig:betaShear}c,d), we set for AND fitness, $\beta = 0.12$ and for OR fitness, $\beta = 0.17$. For a Hagen-Poiseuille flow in a two-dimensional pipe, the flow rate and shear rate are given by,
\begin{align*}
    \mathbf{v} = v_{\text{max}} \left( 1 - \frac{y^2}{H^2} \right) \hat{\mathbf{x}}, \quad \left| \frac{\mathrm{d}v}{\mathrm{d}y} \right| = \frac{2 v_{\text{max}} y}{H^2}. 
\end{align*}
The flow pattern is in the $\hat{x}$ direction and maximal at the center of the pipe, corresponding to $y = 0$. Because of no-slip boundary conditions, flow is zero at the boundaries of the pipe $y = \pm H$ (Fig. \ref{fig:pipe-vortex}a). The shear rate magnitude is given by taking the derivative of the flow rate with respect to $y$ and varies linearly with distance $y$. The shear rate is zero at the center of the pipe and maximal at the boundaries of the pipe.

From our results with a constant shear (Fig. \ref{fig:betaShear}c,d), we expect higher shear regions of the pipe to be occupied by specialists and lower shear regions to be occupied by generalists. However, we see the opposite to occur (Fig. \ref{fig:pipe-vortex}b,c). This is due to boundary and second order effects. Generalist groups on the boundary fragment more often and are able to prevent takeover by mutations. Longer groups are formed in regions of intermediate shear and generate more mutations, leading to a predominately specialist population in this region (Fig. \ref{fig:pipe-vortex}b,c, Supplementary Video 7). 
The fragmenting generalist groups act as a source for specialists groups in the intermediate regions of the pipe. Near the center of the pipe where the shear rate is low, groups do not fragment as quickly and are taken over by cheaters. We therefore see a coexistence of group types across the pipes, with generalists at the boundary, followed by specialists in the intermediate regions (Fig. \ref{fig:pipe-vortex}b,c), and an extinct population due to groups being destroyed by cheaters at the center (Fig. \ref{fig:pipe-vortex}b).

Next we study evolution in a Rankine vortex. The flow and shear profiles for a Rankine vortex with radius $R$ and circulation $\Gamma$ are given by,
\begin{align*}
    \mathbf{v} = 
    \begin{cases} 
        \frac{\Gamma r}{2 \pi R^2} \hat{\theta}, & r \leq R \\
        \frac{\Gamma}{2 \pi r} \hat{\theta}, & r > R
    \end{cases} \quad
    \sigma = 
    \begin{cases}
        0, & r \leq R \\
        \frac{\Gamma}{2 \pi r^2} \hat{\theta}, & r > R .
    \end{cases}
\end{align*}
 The flow pattern is now in the angular direction $\hat{\theta}$. The magnitude of flow increases linearly up to the vortex radius $R$ and then drops as $1/r$, where $r = \sqrt{x^2 + y^2}$ is the distance from the vortex center (Fig. \ref{fig:pipe-vortex}d). The circulation parameter $\Gamma$ corresponds to the line integral of the flow field along a closed path and has units of velocity times length. Here we use it to tune the rate of flow and shear rate. The shear rate $\sigma$ is in the radial direction. It is zero within the vortex $r < R$, maximal at the vortex radius $r=R$, and decreases as $1/r^2$ for $r > R$. There is no shear in the radial direction $\hat{r}$.

The distribution of specialists and generalists in the vortex agrees better with previous results from constant shear (Fig. \ref{fig:pipe-vortex}e,f). We see generalists persist in regions of low shear and specialists mainly reside in an annular region where shear is large (Fig. \ref{fig:pipe-vortex}e,f, Supplementary Video 8). 
In either case we see coexistence of communities with different interaction structures across the full domain. A varying shear profile can therefore allow for different group types to dominate different regions in the fluid, and stably coexist in other regions.

\onecolumngrid
\begin{tcolorbox}[enhanced,width=\textwidth]
\label{box:summary}
{\bf Specialization:}\\
\emph{Large waste diffusion:} Larger waste diffusion lowers self-competition and allows specialists to form denser groups to better utilize public goods secreted by neighbors. \\
\emph{Large public good benefit:} A high benefit for public goods allows specialists to still be fit without secreting as many public goods. This also helps cheaters exploit producers.\\
\emph{Lower secretion costs:} A lower secretion cost can help specialists dominate over generalists, since a smaller penalty for cooperation can make generalists groups too large and more vulnerable to specialist mutations. In this case, large generalist structures are easily taken over by specialist mutants. \\
\emph{Group structure:} Specialists form groups when waste diffusion is larger than public good diffusion and when costs are not too low. When specialists do not form groups, they are easily taken over with cheaters, leading to either ``chasing cheaters,'' (Supplementary Video 6), or extinction. When generalists form smaller, fragmenting groups, they are able to escape take-over by specialists and out-compete specialists.\\
\emph{Fitness type:} The fitness type dictates which types of specialists structure we see -- pure or mixed. In the OR case, specialists generally evolve into structures of isolated types of specialists (Supplementary Video 4). The AND structure is therefore essential to have true division of labor, where each type of specialist exists equally in the group (Supplementary Video 3). \\
{\bf Cheater coexistence:}\\
 \emph{Lack of group structure and small invasion fitness:} Cheaters cannot exist on their own, but must ``predate'' on producers -- generalists or specialists. When producers are fit, and do not form groups, they can grow quicker than cheaters fully taking over. This occurs when waste diffusion is large, and when secretion costs are low. Low secretion cost also lowers the invasion fitness of cheaters, since the advantage of not secreting is lower, helping them to coexist (Supplementary Video 6). \\
{\bf Extinction:}\\
\emph{When cheaters take over:} When the mutation rate and invasion fitness of cheaters is large enough such that they take over groups faster than the groups fragment. This happens when public good benefit or waste diffusion is large, we see this in the top-middle regions of plots in Fig. \ref{fig:pieGraphs}.\\
\emph{When groups are not stable:} When costs are large and public good benefit is low, cooperators need to form denser groups to increase fitness. However, with low waste diffusion, denser groups over-pollute themselves and are no longer stable. We see this in the  bottom-right regions of plots in Fig. \ref{fig:pieGraphs}. \\
{\bf Fluid shear:}\\
\emph{Enhanced group fragmentation:} A shearing flow stretches and distorts groups. It can help groups fragment and reproduce quicker, allowing stability over cheating mutations \citep{uppal2018}.\\
\emph{Enhanced specialization in linear and vortex flows:} Shearing flow can help specialist groups fragment quicker than generalist groups, and therefore transition a population to contain more specialists (Fig. \ref{fig:betaShear}c,d, Supplementary Video 2).  \\ 
\emph{Coexistence of group types:} The local shear rate can determine what groups are stable. A spatially varying flow profile can then allow for coexistence of different community structures across the full fluid domain (Fig. \ref{fig:pipe-vortex}, Supplementary Videos 7 and 8). 
\end{tcolorbox}
\twocolumngrid
\section*{Discussion}

\cite{fletcher2009simple} show that altruism is favored when cooperators are more likely to interact with other cooperators and less likely to encounter cheaters. Such assortment can be attained when populations are viscous \citep{taylor1992altruism} and spatially self structured \citep{stump2018local, wakano2009}. Kin selection is then the main driving evolutionary force of cooperation in spatially structured populations \citep{lion2008self}. Our findings are consistent with these ideas.

More specifically, we have seen that invasion fitness alone does not govern the evolution of interactions within a community. Rather, physical dynamics governing the habitat and the microbes prove highly influential in whether specialized cooperation, generalized cooperation or cheating strategies will dominate, as well as whether multiple types of groups will coexist. We showed that the spatial structure and dynamical properties of communities, as modulated by diffusion constants, decay rates, fluid dynamical forces and domain geometry, can outweigh the role of fitness economics. These physical factors give generalist cooperators groups a fighting chance against specialist cooperators; and generalist and specialist cooperators against cheaters. As such, we view division of labor as a mechanical phenomena as much as an economical one.

While analyzing the competition between different interaction strategies within a community, we also investigated the competition between different kinds of communities. While a given niche with given physical parameters will be typically exclusively dominated by either generalist groups, specialist groups, or cheaters, we also found that for a range of parameters, the physical and economical factors will counteract in a balanced way, leading to the coexistence of multiple interaction structures within one fluid niche. 

A shearing flow can influence the evolution of cooperation in microbial populations \citep{nadell2017flow, uppal2018}. Here we also saw that fluid flow can alter the spatial structure and dynamic properties of communities, and hence the evolution of their cooperative interactions. A shearing flow increases the group size of generalists and fragmentation rates of specialists, and therefore alter the evolutionary stability of the community interaction structure. When the fluid shear profile varies over space, we observe that generalists and specialists not only find the most suitable position for themselves in the fluid and dominate there, they can also coexist in certain regions.  

Many authors view undifferentiated multicellularity as a prerequisite for specialization \citep{pfeiffer2003evolutionary, gavrilets2010rapid, bonner1998origins, rossetti2010evolutionary, michod2007evolution}. In the case where generalists form a spatially homogeneous population and specialists form groups, we have seen that a transition to specialization can split the population into discrete subpopulations, i.e. functional multicellular groups. 
In this light, division of labor can be viewed as a first cause of multicellularity, rather than a consequence. 
 


Though we paid close attention to physical realism, we also made important simplifying assumptions in our first-principles model. First, we assumed identical mutation rates between all pairs of phenotypes, whereas in reality, loss of function mutations are often more likely. Second, for most simulations we took the diffusion constants and decay rates of the two public goods to be identical. Studying cases where $d_1 \neq d_2$ or $\lambda_1 \neq \lambda_2$ could give additional interesting results that we have not explored here. Specifically, we think that the existence of a diffusion length blurs the distinction between public and private goods, and communities might end up with larger numbers of producers of the less diffusive (more private) good and larger numbers of exploiters of the more diffusive (more public) good. We also neglect the finite sizes and complex shapes of microbes, and instead take them as point particles. Additionally, since microbes live in a low Reynolds number environment, we ignore the inertia of microbes, whereas in reality, microbes will themselves influence the fluid flow patterns. This effect will become especially important in highly dense populations and when microbes actively stick to one another or integrate via extracellular polymers. Finally, we neglect the taxis of microbes. In reality, microbes can exhibit complex swimming patterns and move towards or against chemical gradients. 

Theoretical and experimental investigations of these additional factors will provide further insights into the interplay between mechanical factors and evolution of community interactions. 


\section*{Acknowledgments}
This material is based upon work supported by the Defense Advanced Research Projects Agency under Contract No. HR0011-16-C-0062 and National Science Foundation grant CBET-1805157

\section*{Supplementary files}
\begin{description}
    \item[Supplementary Video 1:] Video of generalist population outcompeting specialists and avoiding takeover in the absence of flow in AND fitness type. Parameter values correspond to figure \ref{fig:betaShear}a with cost $\beta = 0.12$, i.e. $d_1 = d_2 = 5 \times 10^{-6} \ \mathrm{cm}^2 \ \mathrm{s}^{-1}, d_w = 15 \times 10^{-6} \ \mathrm{cm}^2 \ \mathrm{s}^{-1}, a_{12} = 6.5 \times 10^{-3} \ \mathrm{s}^{-1}$ and others as given in Table \ref{tab:parameters}.
    \item[Supplementary Video 2:] Video of a shearing Couette flow inducing a transition of a coexisting population to purely specialist in AND case. Parameter values are the same as for video 1 but with Couette flow parameters $H = 20 \ \mathrm{mm}, v_{\text{max}} = 100 \ \mathrm{mm} \ \mathrm{s}^{-1}$.
    \item[Supplementary Video 3:] Specialists in AND case with no flow. Parameter values are $d_1 = d_2 = 5 \times 10^{-6} \ \mathrm{cm}^2 \ \mathrm{s}^{-1}, d_w = 15 \times 10^{-6} \ \mathrm{cm}^2 \ \mathrm{s}^{-1}, a_{12} = 6.5 \times 10^{-3} \ \mathrm{s}^{-1}, \beta = 0.08$ and others as given in Table \ref{tab:parameters}. 
    \item[Supplementary Video 4:] Specialists in OR case with no flow. Parameter values are $d_1 = d_2 = 25 \times 10^{-6} \ \mathrm{cm}^2 \ \mathrm{s}^{-1}, d_w = 60 \times 10^{-6} \ \mathrm{cm}^2 \ \mathrm{s}^{-1}, a_{12} = 7.0  \times 10^{-3} \ \mathrm{s}^{-1}, a_w = 8.0  \times 10^{-3} \ \mathrm{s}^{-1}, \beta = 0.1, \mu = 5 \times 10^{-8} \ \mathrm{s}^{-1}$.
    \item[Supplementary Video 5:] Coexistence of specialists and generalists in OR case without flow. Parameter values are $d_1 = d_2 = 25 \times 10^{-6} \ \mathrm{cm}^2 \ \mathrm{s}^{-1}, d_w = 40 \times 10^{-6} \ \mathrm{cm}^2 \ \mathrm{s}^{-1}, a_{12} = 6.5 \times 10^{-3} \ \mathrm{s}^{-1}, \beta = 0.14, \mu = 5 \times 10^{-8} \ \mathrm{s}^{-1}$. 
    \item[Supplementary Video 6:] Coexistence of all three types: generalists, specialists, and cheaters, in the absence of flow, in AND fitness type. Parameter values are $d_1 = d_2 = 20  \times 10^{-6} \ \mathrm{cm}^2 \ \mathrm{s}^{-1}, d_w = 60  \times 10^{-6} \ \mathrm{cm}^2 \ \mathrm{s}^{-1}, a_{12} = 7.5 \times 10^{-3} \ \mathrm{s}^{-1}, a_w = 105 \times 10^{-3} \ \mathrm{s}^{-1}, \beta = 0.02, \mu = 2 \times 10^{-7}  \ \mathrm{s}^{-1}$.
    \item[Supplementary Video 7:] Generalist/specialist coexistence in Hagen-Poiseuille pipe flow in AND case. Parameters as in figure \ref{fig:pipe-vortex}b, i.e. $d_1 = d_2 = 5 \times 10^{-6} \ \mathrm{cm}^2 \ \mathrm{s}^{-1}, d_w = 15 \times 10^{-6} \ \mathrm{cm}^2 \ \mathrm{s}^{-1}, a_{12} = 6.5 \times 10^{-3} \ \mathrm{s}^{-1} \beta = 0.12$, flow parameters $H = 100, v_{\text{max}} = 125 \ \mathrm{mm} \ \mathrm{s}^{-1}$ and others as given in Table \ref{tab:parameters}.
    \item[Supplementary Video 8:] Generalist/specialist coexistence in Rankine vortex flow, in AND fitness type. Parameters as in figure \ref{fig:pipe-vortex}e, i.e. $d_1 = d_2 = 5 \times 10^{-6} \ \mathrm{cm}^2 \ \mathrm{s}^{-1}, d_w = 15 \times 10^{-6} \ \mathrm{cm}^2 \ \mathrm{s}^{-1}, a_{12} = 6.5 \times 10^{-3} \ \mathrm{s}^{-1} \beta = 0.12$, vortex flow parameters $R = 20 \ \mathrm{mm}, \Gamma = 4000 \pi \ \mathrm{mm}^2 \ \mathrm{s}^{-1}$ and others as given in Table \ref{tab:parameters}
\end{description}

\bibliography{bibliography}
\bibliographystyle{apalike}

\onecolumngrid
\newpage
\renewcommand{\figurename}{Supplementary Figure}
\setcounter{figure}{0}  
\section*{Supplementary material}

\section{Turing analysis}
\label{app:turing}

To understand how each phenotype competes evolutionarily with the other, we solve for Turing patterns formed in each state, in the absence of mutations  and without flow. By performing a linear Turing analysis on each system, we determine the regions in phase space where patterns form.

We perform a Turing analysis by first finding the homogeneous steady states in the absence of diffusion. We then look for conditions where these states are linearly stable to small perturbations. Next we add diffusion to the linearized system and look for conditions that lead to an instability in the system with respect to spatial perturbations. These unstable perturbations will then grow until non-linear terms become relevant. The unstable modes then correspond to Turing patterns giving stripes or spots. We study the Turing patterns of stable systems of generalists and specialists in both AND and OR fitness forms.

\subsection*{Generalist Turing analysis}
For generalists, we simplify our analysis by turning off flow and mutations, to study the native patterns formed by generalist colonies. We then linearize the following system,
\begin{align*}
&\dot{n}_{3}  = d_b \nabla^2 n_{3} + n_{3} f_3(c_\alpha), \\
&\dot{c}_1  = d_1 \nabla^2 c_1 -\lambda_1 c_1 + s_1 n_{3}, \\
&\dot{c}_2  = d_2 \nabla^2 c_2 -\lambda_2 c_2 + s_2 n_{3}, \\
&\dot{c}_3   = d_w \nabla^2 c_3 -\lambda_w c_3 + s_w n_{3} .
\end{align*}
Next, we make the further simplification setting $d_1 = d_2 = d$, $\lambda_1 = \lambda_2 \equiv \lambda$ and $s_1 = s_2 \equiv s$, treating both public goods symmetrically. With this simplification, we can set $c_1 = c_2 \equiv c_{12}$ , and reduce the system to three equations. We then have, for the fitness functions,
\begin{align*}
&f_3^{(\text{AND})} = a_{12} \frac{c_{12}^2}{c_{12}^2 + k_{12}} - a_w \frac{c_3}{c_3 + k_w} - 2 \beta s , \\
&f_3^{(\text{OR})} = a_{12} \frac{2 c_{12}}{ 2c_{12} + k_{12}} - a_w \frac{c_3}{ c_3 + k_w} - 2 \beta s . 
\end{align*}
We next wish to find the stable homogeneous steady states, in the absence of diffusion. We denote by $\mathbf{g}^* = (n_{3}^*, c_{12}^*, c_3^*)$ the homogeneous steady state solution, given by setting the reaction terms to zero, 
\begin{align*}
n_{3}^* f_3 (c_\alpha^*) = 0 , \quad
s n_{3}^* - \lambda c_{12}^* = 0 , \quad
s_w n_{3}^* - \lambda_w c_3^* = 0.
\end{align*}
The diffusion terms vanish for spatially homogeneous states. The steady state value for the chemicals are then given by,
\begin{align*}
c_{12}^* = \frac{s}{\lambda} n_{3}^*, \quad c_3^* = \frac{s_w}{\lambda_w} n_{3}^*.
\end{align*}
The solution for $n_{3}^*$ depends on the form of the fitness function.
In the AND case, $n_{3}^*$ is given by solving the qubic equation,
\begin{align*}
& \frac{s^2 s_w}{\lambda^2 \lambda_w} ( a_{12} - a_w - 2 \beta s) n_{3}^3 + \frac{k_w s^2}{\lambda^2} (a_{12} - 2 \beta s ) n_{3}^2  - \frac{k_{12} s_w}{\lambda_w} (a_w + 2 \beta s) n_{3}  -2 \beta k_{12} k_w s = 0 .
\end{align*} 
We enforce $a_{12} - a_w - 2 \beta s < 0$ for stability, otherwise the dense state will become unstable as the hill forms become saturated. Since the linear and constant coefficients are also negative, by Descartes rule of signs, in order to have any positive solution, we need $a_{12} - 2 \beta s > 0$. This implies we have up to two positive solutions, since there are two sign changes.


In the OR case, we have $n_{3}^*$ given by solving the quadratic equation,
\begin{align*}
\frac{2 s s_w}{\lambda \lambda_w} ( a_{12} \!-\! a_w \!-\! 2 \beta s) n_{3}^2 + \frac{1}{\lambda \lambda_w} \left[ 2 a_{12} k_w \lambda_w s \!-\! a_w k_{12} \lambda s_w - 2 \beta s( 2 k_w \lambda_w s + k_{12} \lambda s_w) \right] n_{3} - 2 \beta k_{12} k_w s = 0
\end{align*}
For this to have a positive solution, we require
\begin{align*}
2 a_{12} k_w \lambda_w s - a_w k_{12} \lambda s_w - 2 \beta s( 2 k_w \lambda_w s + k_{12} \lambda s_w) > 0 .
\end{align*}

Next we perform a linear stability analysis for each case. We let $\mathbf{g} \equiv (n_{3}, c_{12}, c_3)^T - \mathbf{g}^*$, be a perturbation from the steady state. Our linearized system then looks like,
\begin{align*}
\frac{\partial}{\partial t} \mathbf{g} = A \mathbf{g},
\end{align*}
where our stability matrix $A$ is given as
\begin{align*}
A = \begin{pmatrix} 0 & f_{,12} & f_{,w} \\
s & -\lambda & 0 \\
s_w & 0 & -\lambda_w 
\end{pmatrix} .
\end{align*}
Here $f_{,12}$ and $f_{,w}$, in each case, are given by
\begin{align*}
&f^{(\text{AND})}_{,12} = \left. n_3 \frac{2 a_{12} k_{12} c_{12}}{(c_{12}^2 + k_{12})^2 }\right|_{ n_3^*, c_{12}^*}, \\
&f^{(\text{OR})}_{,12} = \left.  n_3 \frac{2 a_{12} k_{12}}{(2c_{12} + k_{12})^2 }\right|_{ n_3^*, c_{12}^*}, \\
&f^{(\text{AND})}_{,w}  = f^{(\text{OR})}_{,w} = \left.-  n_3 \frac{a_w k_w}{(c_3 + k_w)^2}\right|_{ n_3^*, c_3^*} .
\end{align*}
The characteristic polynomial is given in terms of the invariants of a $3 \times 3$ matrix,
\begin{align*}
\Lambda^3 - \mathrm{tr}(A) \Lambda^2 + \frac{1}{2} ((\mathrm{tr}(A))^2 - \mathrm{tr}(A^2)) \Lambda - \mathrm{det}(A) = 0
\end{align*}
Again, using Descartes rule of signs, we have the following requirements for linear stability, and with the knowledge that $\mathrm{tr}(A)$ is negative, we require
\begin{align*}
&\mathrm{det}(A)  < 0 \\
&(\mathrm{tr}(A))^2 - \mathrm{tr}(A^2)  > 0 .
\end{align*}
Next we include diffusion and look for a Turing instability. We expand our solution in terms of Fourier modes with exponential growth,
\begin{align*}
\mathbf{g}(\mathbf{x},t) = \sum_k c_k e^{i \mathbf{k} \cdot \mathbf{x}} e^{\Lambda (\mathbf{k}) t}.
\end{align*}
Plugging this into our linearized system, we get the eigenvalue equation, $(-k^2 D + A) \mathbf{g} = \Lambda \mathbf{g}$. Where, 
\begin{align*}
D = \begin{pmatrix} d_b & 0 & 0 \\
0 & d & 0 \\
0 & 0 & d_w 
\end{pmatrix} .
\end{align*}
If we denote by $M(k) = -k^2 D + A$, the characteristic equation for this system is now given as
\begin{align*}
\Lambda(k)^3 - \mathrm{tr}(M(k)) \Lambda^2 + \frac{1}{2} \big{[} (\mathrm{tr}(M(k)))^2 - \mathrm{tr}(M(k)^2) \big{]} \Lambda(k) - \mathrm{det}(M(k)) = 0
\end{align*}
We now use Descartes rule of signs once again, this time to get an instability. First, we have $\mathrm{tr}(M) = -k^2 \mathrm{tr}(D) + \mathrm{tr}(A) < 0$, which makes the quadratic term positive. For the linear term, we have
\begin{align*}
(\mathrm{tr} &(M(k)))^2 - \mathrm{tr}(M(k)^2) = [ -k^2 \mathrm{tr}(D) + \mathrm{tr}(A)]^2 - \mathrm{tr}([-k^2 D + A]^2) \\
& = k^4 (\mathrm{tr}(D))^2 - 2 k^2 \mathrm{tr}(D) \mathrm{tr}(A) + (\mathrm{tr}(A))^2  - \left[ k^4 \mathrm{tr}(D^2) - k^2 \mathrm{tr}(DA) -k^2 \mathrm{tr}(AD) + \mathrm{tr}(A)^2 \right] \\
& = k^4 [(\mathrm{tr}(D))^2 - \mathrm{tr}(D^2)] + (\mathrm{tr}(A))^2 - \mathrm{tr}(A^2)  + k^2[ \mathrm{tr}(DA) + \mathrm{tr}(AD) - 2 \mathrm{tr}(D) \mathrm{tr}(A)] .
\end{align*}
Now since $(\mathrm{tr}(D))^2 - \mathrm{tr}(D^2) = (d_b + d + d_w)^2 -  (d_b^2 + d^2 + d_w^2) > 0$, \ $\mathrm{tr}(DA) + \mathrm{tr}(AD) - 2 \mathrm{tr}(D) \mathrm{tr}(A) = 2(d_w \lambda +  d \lambda_w + d_b ( \lambda + \lambda_w )) > 0$, and we require $(\mathrm{tr}(A))^2 - \mathrm{tr}(A^2) > 0$ from linear stability, the linear term is also positive. Therefore, by Descartes rule of signs, the requirement for Turing instability reduces to,
\begin{align*}
\mathrm{det}(-k^2 D + A) > 0 ,
\end{align*}
for some range of $k > 0$. 

Inspecting this further, we see that $\mathrm{det}(-k^2 D + A)$ gives a cubic polynomial in $k^2$ which goes to negative infinity as $k \rightarrow \infty$ and passes through $\mathrm{det} (A) < 0$ at $k = 0$. We therefore see that a Turing instability occurs when the local maximum of $\Psi(k^2) = \mathrm{det}(-k^2 D + A) \geq 0$. We therefore have that the critical wavenumber is given by maximizing  $\Psi(k^2)$. Let $w = k^2$, then 
\begin{align*}
\Psi (w) &= - \mathrm{det}(D) w^3 - \left( d_b d_w \lambda + d_b d \lambda_w \right) w^2 \\
& + \left(d f_{,w} s_w - d_b \lambda \lambda_w + d_w f_{,12} s \right)w + \mathrm{det}(A) .
\end{align*}
The local maximum is then given by taking $\frac{\mathrm{d} \Psi}{\mathrm{d} w} = 0$ and taking the positive root,
\begin{align*}
w^* = \frac{-b - \sqrt{b^2 - 4ac}}{2a},
\end{align*}
where $a = -3 \mathrm{det}(D)$, $b = -2(d_b d_w \lambda + d_b d \lambda_w)$, $c = (d f_{,w} s_w - d_b \lambda \lambda_w + d_w f_{,12} s ) $. Given the critical value $w^*$ and a public good diffusion $d$, we can find the critical waste diffusion that gives a Turing instability by setting $\Psi(w^*,  d^c_w) = 0$ and solving for  $d^c_w$. This then gives a relation for  $d^c_w$ as a function of the other system parameters, when it exists. The theoretical region of Turing instability is compared against numerical results in Fig. \ref{fig:turing}.

\subsection*{Specialist Turing analysis}
The specialist case can be treated in a similar manner. We again look at dynamics without flow or mutations to study the native colony patterns formed by specialists. Here we take $n_{3} = 0$ and because of symmetry, take $n_1 = n_2 \equiv n_s$. We therefore again reduce our system to three equations and get the following system,
\begin{align*}
\dot{n}_{s} & = d_b \nabla^2 n_{s} + n_{s} f(c_\alpha) \\
\dot{c}_{12} & = d \nabla^2 c_{12} -\lambda c_{12} + s n_{s} \\
\dot{c}_3  & = d_w \nabla^2 c_3 -\lambda_w c_3 + 2 s_w n_{s} 
\end{align*}
The factor of $2$ in the waste chemical term comes from the fact that both populations $n_1$ and $n_2$ secrete waste while only one secretes the public good. We also have that the factor of 2 in front of the cost in the fitness functions now becomes a one. We therefore have the same results as in the generalist case with the substitutions,
\begin{align*}
n_{3} \rightarrow n_s, \quad 2 \beta \rightarrow \beta, \quad s_w \rightarrow 2 s_w.
\end{align*}
With these substitutions, the results from the generalists case applies also to the specialists. In general, the Turing pattern forming region will be different for the two states. These differing Turing regions and pattern types effect what type of state, generalist or specialist, will be more stable evolutionarily when also considering mutations. 


 \section{Model details and parameter sensitivity}
\label{app:para}
\subsection*{Model details}
We provide additional details on the implementation of our agent based stochastic simulations performed in two dimensional space. Our time step $\Delta t$ is chosen to be $\Delta t = 1.0\ \mathrm{s}$. Microbes diffuse and advect via a biased random walk scheme. Specifically, a step $\delta$ is chosen as $\delta = \sqrt{4 d_b \Delta t}$ and is in direction either up, down, left, or right with uniform random probability. We then add a bias given by the flow velocity $ \vec{v} \Delta t$. We note $\Delta t$ is sufficiently small for this to give an accurate representation of the diffusion-advection process with our given diffusion and flow parameters. Microbes also secrete chemicals locally onto a discrete grid that then diffuse and advect using a finite difference scheme. Diffusion and secretion rates are given in Table \ref{tab:parameters} and in figure captions when they differ. Secretion is given by adding $s_\alpha \Delta t$ for each chemical $\alpha$ to the cell occupied by the microbe. We use an explicit forward Euler method using central difference for diffusion and a first order upwinding scheme for advection terms. Spatial discretization and time step are chosen so that the scheme converges and no difference was seen using smaller discretizations. For reproduction and death steps, we compute a probability dependent on the local fitness of a microbe and time step, given by $f(\mathbf{c}) \Delta t$. Note $\Delta t$ and fitness constants are always sufficiently small so that $|f \Delta t| \ll 1$. If $f \Delta t$ is negative, the microbes die with probability 1, if $f \Delta t$ is between 0 and 1 they reproduce an identical offspring with probability $f \Delta t$. Upon reproduction, offspring are placed at the same location as their parent. Numerical simulations were performed by implementing the model described using Matlab. The source code for discrete simulations is provided as a supplemental file and videos of these simulations are provided as Supplementary Video files.

\subsection*{System parameters}

Our model contains parameters for chemical benefit or harm, secretion cost, diffusion, decay rates, secretion rates, flow rates, and mutation rates. Chemical concentration values can be rescaled to remove saturation constants. Therefore, saturation constants do not play a major role. 

We note that fitness constants of chemical effects and cost are constrained by the relation in our Turing analysis (Supplementary Section \ref{app:turing}), giving $a_{12} < a_w$ and $a_{12} > 2 \beta s$. We fix $a_{w}$ throughout for both AND and OR fitness cases. This value establishes a time scale for microbe growth and death. The largest growth rate in our model is typically on the order of a division every $10^{-3} \ \mathrm{s}^{-1}$ which is around empirical values of once every 20 to 30 minutes at most \citep{gibson2018distribution}. Values of $\alpha_{12}$ and $\beta$ are then varied across their valid range as given by the above constraints (e.g. Fig. \ref{fig:pieGraphs}). Thus by varying fitness, we cover a range of relevant growth rates.

Diffusion constants also correspond to empirical values typically on the order of $10^{-6} \ \mathrm{cm}^2 \ \mathrm{s}^{-1}$ and can vary over a couple orders of magnitude \citep{Ma2005,Kim1996}. Diffusion parameters are also varied to give various Turing patterns. The eco-evolutionary dynamics observed in the paper are strongly due to spatial effects and depend on how patterns change across other parameters. We therefore chose diffusion constants that gave sensitive dependence on cost and benefit parameters. We also varied waste diffusion along with fitness constants (Fig. \ref{fig:pieGraphs}) to cover a range of relevant growth rates together with different emergent group structures. 

\begin{figure}
    \centering
    \includegraphics[width=0.95\textwidth]{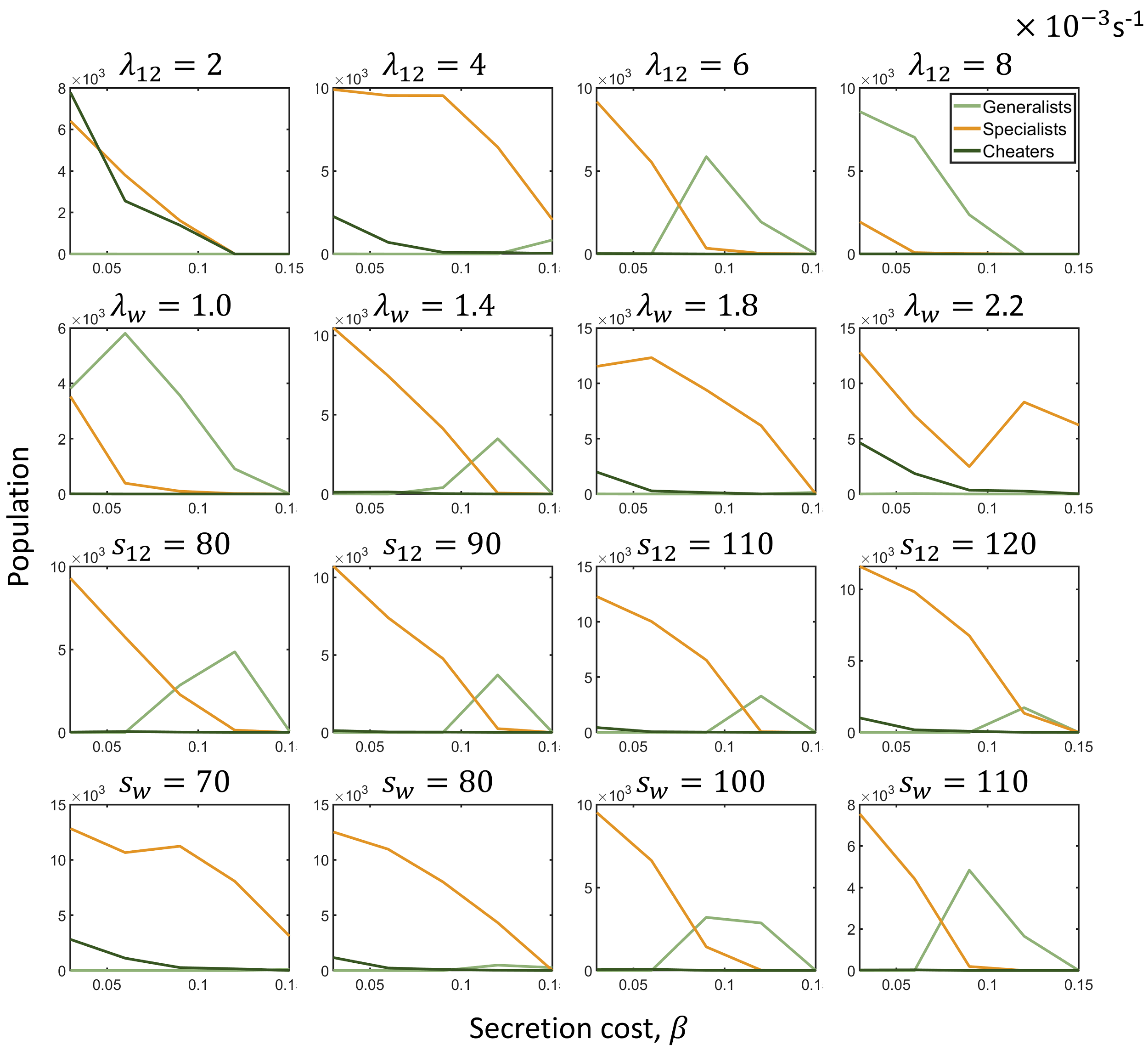}
    \caption{{\bf Effects of varying decay constants and secretion rates} In the top row, we vary the public good decay constant above and below values used throughout the paper. The effect of increasing (decreasing) the decay constant corresponds to decreasing (increasing) the steady state public good concentration. This can be related to adjusting the public good constant $a_{12}$. Lower decay rates coincide with higher public good benefit $a_{12}$. In the second row, we vary the waste decay constant. Similarly, varying the waste decay constant adjusts the steady state waste concentration. This is somewhat captured by varying the waste diffusion constant or again the public good benefit. In the third row we vary the public good secretion rate. Varying the public good secretion interesting has little effect when looked at against secretion cost $\beta$. This is because higher public good secretion will increase the public good concentration, and hence fitness, but will also increase the cost. These competing effects somewhat cancel out and the effect of varying $s_{12}$ is diminished. In the last row we vary the waste secretion rate. Varying the waste secretion rate effects the steady state concentration value of the waste chemical and has similar effect as varying the waste decay constant. A higher waste secretion rate can be also seen as a lower waste decay rate. Parameter values for diffusion are $d_1 = d_2 = 5 \times 10^{-6} \ \mathrm{cm}^2 \ \mathrm{s}^{-1}, d_w = 15 \times 10^{-6} \ \mathrm{cm}^2 \ \mathrm{s}^{-1}$, for public good benefit, $a_{12} = 6.5\times 10^{-3} \ \mathrm{s}^{-1}$. Flow rate is set to 0. The rest of the parameter values are given in Table \ref{tab:parameters}.} 
    \label{fig:lambda_sec}
\end{figure}

We do not vary decay or secretion constants in the paper, however the effects of decay and secretion can be accounted for by rescaling the chemical concentrations. This is somewhat captured by varying fitness constants. We see in Fig. \ref{fig:lambda_sec} that varying decay and secretion rates can be understood much in the same way as varying public good benefit and/or waste diffusion (Fig. \ref{fig:pieGraphs}). Increasing (decreasing) the public good decay rate decreases (increases) the steady state public good concentration. We can capture this effect by increasing (decreasing) the public good benefit $a_{12}$ for lower (higher) decay rates. We do see larger public good decay selects for more generalists and smaller public good decay selects for more cheaters (Fig. \ref{fig:lambda_sec}). Similarly, the waste decay rate effects the steady state concentration for waste chemicals. This is also somewhat captured by varying the public good benefit constant relative to the harm from waste. We can therefore compare increasing (lowering) the waste decay rate to increasing (lowering) the public good benefit (Fig. \ref{fig:pieGraphs}). This is not an exact mapping, but the steady state behavior can be qualitatively understood this way. 

Varying the public good secretion rate has little effect. This is due to competing effects. Increasing the public good secretion rate increases the public good concentration but also increases the cost paid by microbes. Varying the waste secretion concentration changes the steady state concentration of the waste chemical. This can be understood in much the same way as varying the waste decay rate. A higher waste secretion rate effectively corresponds to a lower waste decay rate and vice versa (Fig. \ref{fig:lambda_sec}). We therefore see the effects of decay and secretion concentrations can effectively be taken into account by varying fitness constants as done in the paper (Fig. \ref{fig:pieGraphs}). Decay and secretion constants throughout the paper were picked to give stable Turing patterns which then vary with other parameters.

Mutation and flow rates are also picked around empirical values \citep{drake1998rates, rusconi2015microbes}. Flow rates were varied along fitness and diffusion parameters where they had the largest observed effect (Fig. \ref{fig:betaShear} and \ref{fig:pipe-vortex}) Cost and benefit parameters were also fixed where the effect of flow was strongest. 

Lastly, we note that we treat chemicals $c_1$ and $c_2$ equally for simplicity. Treating quantities such as $d_1 \neq d_2$ or $\lambda_1 \neq \lambda_2$ could give additional interesting results that we have not explored here. The diffusion-decay length gives a continuous distinction of public vs private goods. A shorter diffusion-decay length corresponds to a more local public good that is less susceptible to exploitation. Kin selection may then cause exploitation of more diffusive public goods and the relative ratio of producers will shift to have more producers of the local good relative to the more diffusive good in specialist groups.


\section{Effective group models}
\label{app:effectiveGroup}
To get a better understanding of the results in Fig. \ref{fig:betaShear}, we developed an effective model of ordinary differential equations describing the dynamics of each type of group. We assume that each phenotype, generalists and specialists, form groups that grow and reproduce with different rates. The growth of the groups are given by a logistic equation, with a carrying capacity dependent on the area of the system. 
In order to fully describe the logistic growth, we also need to include transient states that are not stable but are continuously formed. The transient states therefore restrict the space in which the stable states are allowed to grow. We denote generalists groups by $G$. Mutations typically cause groups to go down in number of secreted compounds. Back-mutants do not fixate in groups since they are paying more cost than their neighbors, so we neglect transitions towards secreting more goods. Groups composed of generalists and specialists are therefore transitioning to a specialist state and are denoted by $T$. Mixed specialist groups made of microbes that secrete different public goods are represented by $M$. Pure specialists groups composed of microbes secreting only one public good are denoted by $P$. Pure specialist groups only occur in the OR case, and we exclude them in our effective model for the AND case. Finally, groups with cheaters that secrete no public good are denoted by $C$.  The evolutionary paths the groups can take are illustrated in Fig. \ref{fig:model_schematic}c. 


\subsection*{Effective model for AND type fitness}

The group dynamics for the AND case are described by the following set of equations,
\begin{align}
&\dot{G} = r_g \left[1 - \Sigma/k \right] G - (1/2) m_g \mu G, \label{eq:gen} \\
&\dot{T} = - r_t T + (1/2) m_g \mu G - (1/2) m_t \mu T, \\
&\dot{M} = r_m \left[1 - \Sigma/k \right] M - (1/2) m_m \mu M +  (1/2) m_t \mu T, \\
&\dot{C} = - r_c C +  (1/2) m_m \mu M,
\label{eq:mut}
\end{align}
where the constants $m_i$ correspond to the number of microbes in a group of type $i$, and $\Sigma=m_g G + m_t T + m_m M + m_c C$ is the total population. For transient groups, the constant $m_i$ gives the average group size over its lifetime. For stable groups, we take an average over many groups after they have stabilized. The rates $r_i$ give the fragmentation rates for stable groups and the extinction rates for transient groups. Transient groups may also occasionally fragment, especially at low secretion costs, but eventually will go extinct. The extinction rates $r_i$ for transient groups then gives the net fragmentation/extinction rate, and is treated as an effective extinction rate. The carrying capacity for the system is given by $k$. Mutations cause groups to go down in number of secreted compounds (Fig. \ref{fig:model_schematic}c), and are given by terms proportional to the mutation rate $\mu$.  

\begin{figure}
    \centering
    \includegraphics[width=0.8\textwidth]{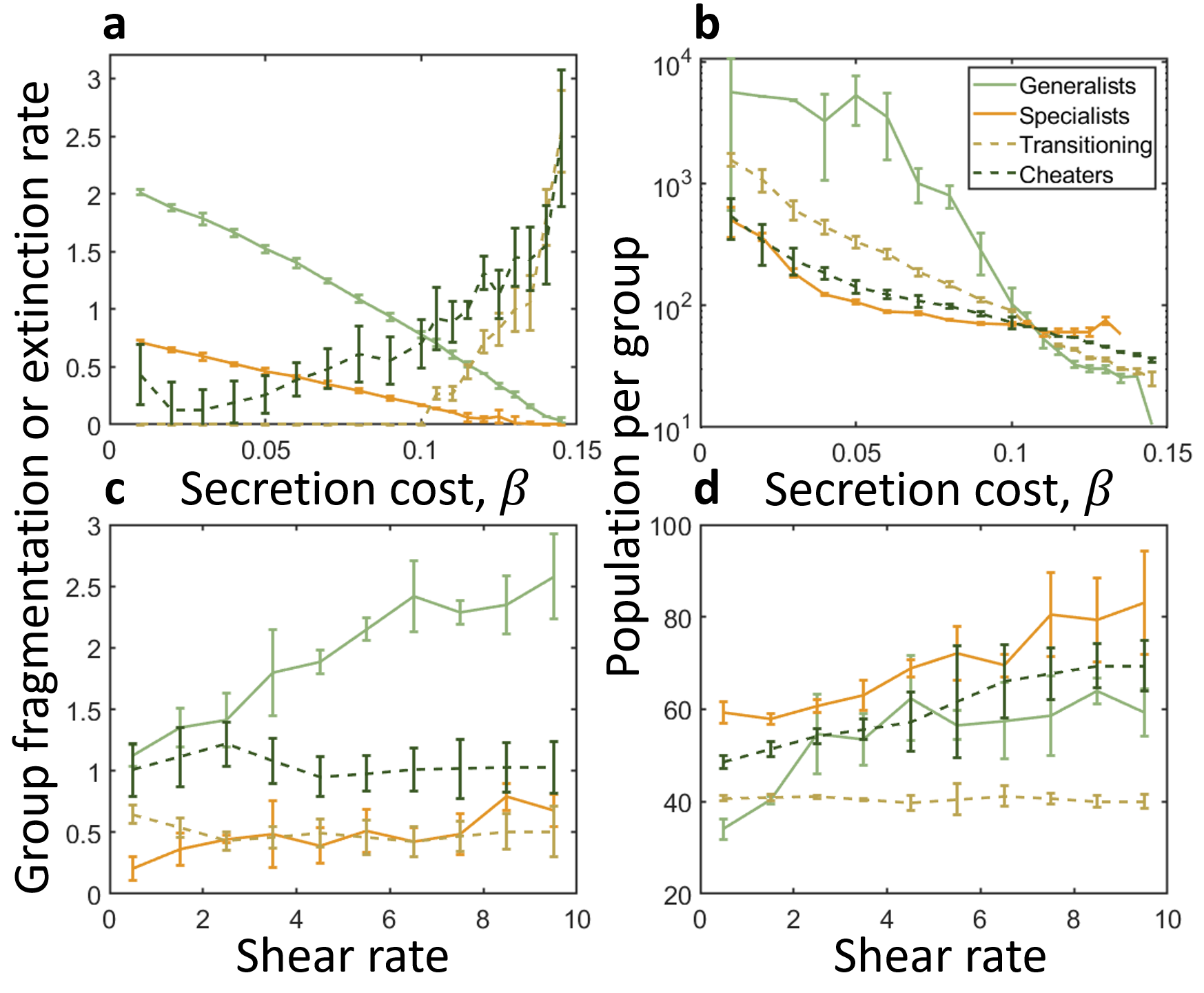}
    \caption{{\bf Group fragmentation or extinction rates and populations used in effective model for AND fitness variant} Group fragmentation (for stable groups) and extinction rates (for transitioning and cheater groups) were measured empirically from simulations seeded with just one type of group and without mutations to observe their native dynamics. {\bf a,b} How parameters vary with secretion cost. We see that generalist groups usually fragment quicker than specialist groups, however at around secretion cost $\beta = 0.10$, specialist groups become larger than generalist groups. They are then more susceptible to cheating mutations than generalists, and generalist groups become evolutionarily favorable. Extinction rates of transient groups increase with cost since the invasion fitness of the specialist (in the case of transitioning groups) or cheater (for cheating groups) is larger compared to producers as cost increases. {\bf c,d} How parameters vary with shearing flow. As we increase shear rate, generalists groups increase in size quicker than specialists. The larger generalist groups then generate more specialist mutations. At around a shear rate of 3, we see in Fig. \ref{fig:betaShear} that specialists take over and generalists are no longer stable. Here, because of the logistic growth, at high density, all groups will reproduce slower and mutations will cause generalist groups to become specialists. Specialist groups will also get taken over by cheaters, which then go extinct. As the population density decreases, specialist groups will be able to fragment quicker. Since back mutations do not occur, once the generalist population is lost, they do not recover. Thus, even though generalist groups reproduce faster at low densities, evolutionarily, specialists become more stable at high shear.}
    \label{fig:effective_and}
\end{figure}

The factor of 1/2 in front of mutation terms is due to mutations not always fixing in a group before it splits in two. As a mutation begins to fixate, the group may fragment in two, giving an effective success rate of roughly 1/2 the time. The exact fixation probability can be added as an extra measured parameter, but we keep it as 1/2 for the sake of simplicity.
Parameter values were obtained from simulations without mutations to determine the natural state and growth of an isolated phenotype. For generalist and specialists, we seeded our simulations with an initial group and let it grow to fill the full simulation area. We then fit a logistic curve to the population over time to get the growth rates and carrying capacities of the population. To get the average population per group, we take the total population at the end of a simulation and divide by the number of groups. For transient groups we measure the extinction rate and average population of the group over its lifetime. For transitioning groups we seed our simulations with a generalist group and a single specialist mutation. We then measure the time it takes for extinction and average the population over the lifetime of the group. Similarly, for the cheating group we start with a specialist group with a single cheating mutation. 

By setting (Equation \ref{eq:gen} -Equation \ref{eq:mut}) to zero, we obtain the steady state values for each type of group. The possible steady states are the trivial extinct state, where all group populations vanish, a stable state of mixed specialists and cheaters, and a stable state where all phenotypes coexist. The specialist/cheater state is given by,
\begin{align*}
M^* = \frac{k r_c (2  r_m - \mu  m_m )}{\mu m_c m_m r_m + 2 m_m r_c r_m} \quad
C^* = \frac{k \mu (2 r_m -  \mu m_m)}{2 \mu  m_c r_m  +4 r_c r_m}
\end{align*}
In the stable state where all group types coexist, the number of generalists groups is given by,
\begin{align*}
G^* = \frac{k r_c ( 2 r_g - \mu  m_g) (\mu  m_t  + 2 r_t) (m_m r_g - m_g r_m)}{ m_g r_g [  2m_m r_c r_g(3 m_t \mu + 2 r_t)  -  4 m_g r_c r_m (\mu m_t + r_t) + \mu ^2 m_c   m_m m_t r_g ]}
\end{align*}
The term in square brackets in the denominator is positive when,
\[r_g > \frac{4 m_g r_c r_m (m_t \mu + r_t)}{m_m (m_c m_t \mu^2 + 6 m_t \mu r_c + 4 r_c r_t)} \]
and since,
\[ \frac{4 m_g r_c r_m (m_t \mu + r_t)}{m_m (m_c m_t \mu^2 + 6 m_t \mu r_c + 4 r_c r_t)} < \frac{4 m_g r_c r_m (m_t \mu + r_t)}{m_m (4 m_t \mu r_c + 4 r_c r_t)} = \frac{m_g r_m}{m_m},\]
for the generalist solution to be positive, it is sufficient for the generalist group fragmentation rate to satisfy,
\begin{align*}
 r_g > \mathrm{max}\left( \frac{\mu m_g}{2}, \frac{m_g r_m}{m_m} \right). 
\end{align*}
Therefore, the generalist groups must fragment (1) faster than mutations arise within generalists groups, and (2) faster than the specialist group fragmentation times the relative size of generalist groups to specialist groups. Hence larger generalist groups also need to fragment at a faster rate and smaller specialist group size ($m_m$) also increases the required fragmentation rate for generalists. The  population of other group types can be given from the ratios,
\begin{align*}
\frac{T^*}{G^*} = \frac{\mu m_g}{\mu  m_t  + 2 r_t}, \quad
\frac{M^*}{G^*} = \frac{\mu  m_g m_t r_g}{(\mu m_t } + 2 r_t) (m_m r_g - m_g r_m), \quad
\frac{C^*}{G^*} = \frac{\mu^2 m_g m_m m_t r_g}{2 r_c (\mu  m_t  + 2 r_t) (m_m r_g - m_g r_m)}.
\end{align*}

We plot the results of our effective model as solid curves against simulation results (Fig. \ref{fig:betaShear}) and get good overall agreement with the numerical simulations. At low costs $(\beta)$, microbes form stripes or become homogeneous, and the groups structure assumption of our effective model breaks down.

\subsection*{Effective model for OR type fitness}

In the OR case, pure specialist groups are now stable. Since we treat both chemicals symmetrically, and since the chemicals enter additively in the fitness function, pure specialist and mixed specialist groups are equivalent.  However, mixed specialist groups are very rarely seen since once a specialist mutation occurs in a generalist group, it quickly sweeps the group and fixates as a pure specialist group, before a complementary specialist arises. Therefore we just label all specialist groups as pure specialists. Furthermore, transition groups are no longer considered. Once a specialist mutation occurs in a generalist group, it is counted as a pure specialist group. This is because the transitioning group will fixate as a pure specialist group, unlike in the previous case where transitioning groups would die out unless a complementary specialist mutation arises within the lifetime of the group.
With these considerations, we build and solve an effective model only including generalists, pure specialists, and cheaters. The effective model is described by the following dynamical equations,  
\begin{align}
\dot{G} &=  r_g \left[1 - \Sigma/k \right] G  - (1/2) m_g \mu G \label{eq:genOR} \\[2mm]
\dot{P} &=  r_p \left[1 - \Sigma/k \right] P - (1/2) m_p \mu P +  (1/2) m_g \mu G \label{eq:pureOR} \\
\dot{C} &= -r_c C +  (1/2) m_p \mu P 
\label{eq:cheatOR}
\end{align}
Here $\Sigma$ is now defined as $\Sigma = m_g G + m_p P + m_c C$. Setting (Equation \ref{eq:genOR} - Equation \ref{eq:cheatOR}) to zero to get steady states, we again get two non-trivial solutions. One with only specialist and cheating groups, and one where all three coexist. For the solution with only specialists and cheaters, we get
\begin{align*}
P = \frac{ k r_c ( 2 r_p - \mu  m_p)}{\mu  m_c m_p r_p + 2 m_p r_c r_p}, \qquad
C = \frac{k \mu ( 2 r_p - \mu m_p ) }{2 \mu  m_c r_p + 4 r_c r_p} .
\end{align*}
For the solution where all three coexist, we get for the generalists,
\begin{align*}
G^* = 
 \frac{k r_c (2 r_g - \mu  m_g ) (m_p r_g - m_g r_p)}{m_g r_g (\mu  m_c m_p r_g + 4 m_p r_c r_g - 2 m_g r_c r_p)} .
\end{align*}
Here the denominator is positive if,
\[r_g > \frac{2 m_g r_c r_p}{m_p( m_c \mu + 4 r_c)},\]
and since,
\[ \frac{2 m_g r_c r_p}{m_p( m_c \mu + 4 r_c)} < \frac{m_g r_p}{m_p},\]
a sufficient condition for the generalist solution to be positive is given by,
\[ r_g > \mathrm{max}\left( \frac{\mu m_g}{2}, \frac{m_g r_p}{m_p} \right).\]
Therefore, like in the AND case, generalist groups must fragment faster than mutations arise within generalists groups, and faster than the specialist group fragmentation times the relative size of generalist groups to specialist groups. The other groups are given by taking the ratios,
\begin{align*}
\frac{P^*}{G^*} = \frac{m_g r_g}{m_p r_g - m_g r_p}, \qquad
\frac{C^*}{G^*} = \frac{\mu  m_g m_p r_g }{2 m_p r_c r_g - 2 m_g r_c r_p } .
\end{align*}
We plot the results of our effective model in the OR case against simulation results (Fig. \ref{fig:betaShear}) and again get a good overall agreement with the numerical simulations.
\begin{figure}
    \centering
    \includegraphics[width=0.8\textwidth]{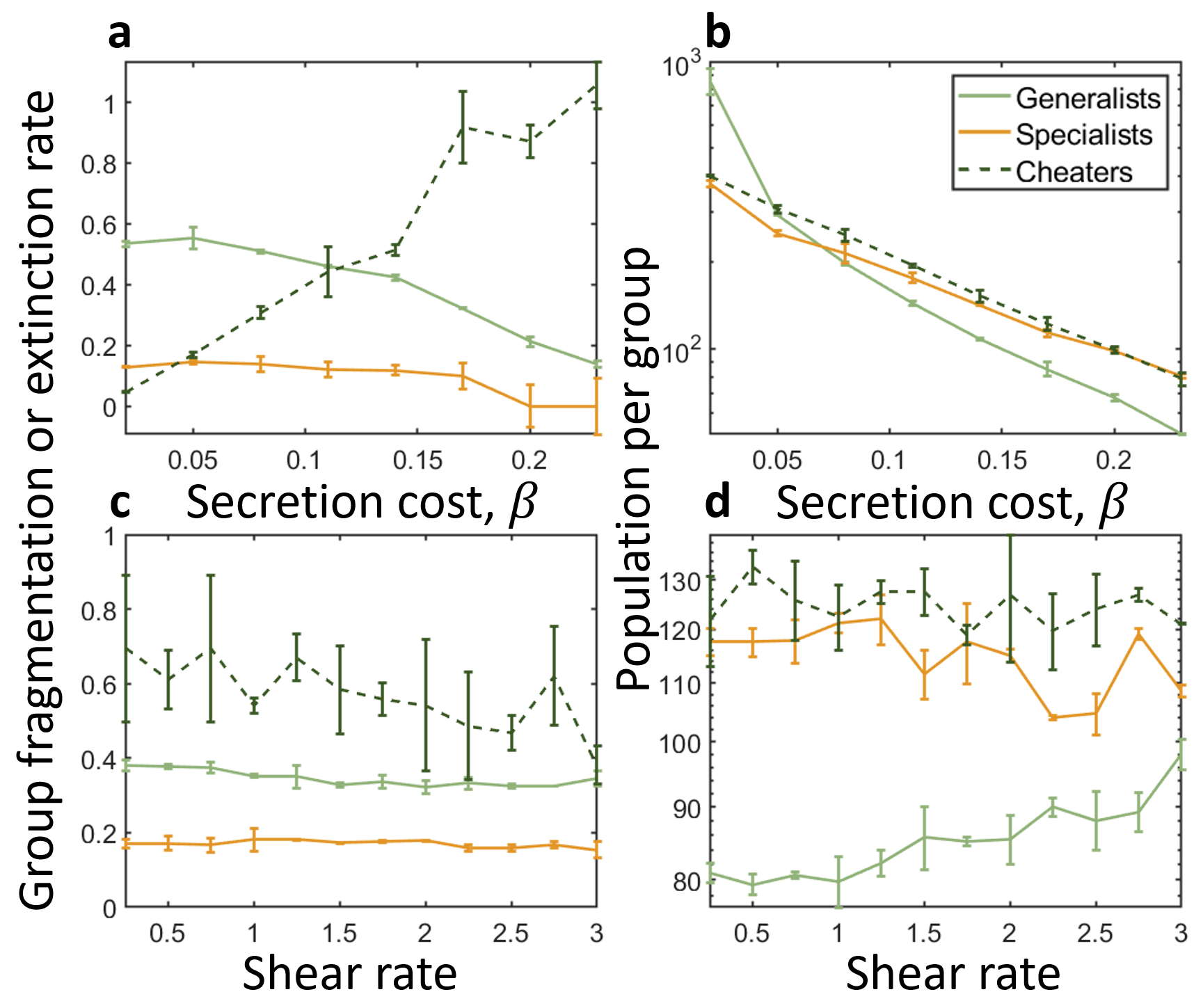}
    \caption{{\bf Group  fragmentation or extinction rates and  populations used in effective model for OR fitness variant} As with the AND case, populations and rates are obtained empirically from simulations without mutations, to observe the native dynamics of each group type. {\bf a,b} How rates and populations vary with secretion cost. {\bf a,} The group  fragmentation rates decrease with secretion cost, but not by a lot. Extinction rates of cheating groups increase with cost since the invasion fitness of the cheater is larger compared to producers at higher costs. {\bf b,} As generalists groups become smaller in size, they become less vulnerable to specialist mutations,  since the smaller groups generate less mutations, and are able to coexist with specialist groups (Fig. \ref{fig:betaShear}). {\bf c,d} How rates and populations vary with shear. {\bf c,} In this case, we see shear does not cause a large difference in the group  fragmentation rates. {\bf d,} Shear causes generalists groups to grow in size relative to specialists. This growth in size makes generalists more susceptible to specialist mutations since larger generalist groups generate more mutations and smaller specialist groups generate less cheating mutations. In the long time limit, specialists take over generalists groups (Fig. \ref{fig:betaShear}).}
    \label{fig:effective_or}
\end{figure}


\end{document}